\documentclass[journal]{IEEEtran}

\usepackage{amsmath,amsfonts,amssymb,amsthm}
\usepackage{algorithmic}
\usepackage{algorithm2e}
\usepackage{array}
\usepackage{textcomp}
\usepackage{stfloats}
\usepackage{url}
\usepackage{verbatim}
\usepackage{graphicx}
\usepackage[loadonly]{enumitem}
\usepackage[T1]{fontenc}
\usepackage{multirow}

\ifCLASSOPTIONcompsoc
\usepackage[caption=false,font=normalsize,labelfont=sf,textfont=sf]{subfig}
\else
 \usepackage[caption=false,font=footnotesize]{subfig}
\fi

\usepackage[backend=biber,style=ieee,natbib=true,citestyle=numeric,doi=false,isbn=false,maxnames=99,mincitenames=1,maxcitenames=2]{biblatex}
\usepackage{xpatch}
\xpatchbibmacro{textcite}{\addspace}{\addnbspace}{}{}
\xpatchbibmacro{Textcite}{\addspace}{\addnbspace}{}{}
\DefineBibliographyStrings{english}{
	andothers = et~al\adddot\addspace
}

\usepackage{color}
\usepackage{bbold}
\usepackage{siunitx}
\usepackage{booktabs}
\usepackage{color,soul}
\usepackage{xcolor}
\usepackage[hidelinks,bookmarks=false,draft]{hyperref}
\usepackage{paralist}
\usepackage{tikz}
\usetikzlibrary{decorations}
\usetikzlibrary{decorations.pathmorphing}
\usetikzlibrary{decorations.pathreplacing}
\usetikzlibrary{shapes}
\usetikzlibrary{through}
\usetikzlibrary{backgrounds,positioning}

\definecolor{viridis_light}{RGB}{253, 231, 37}
\definecolor{viridis_dark}{RGB}{68, 1, 84}

\definecolor{green}{RGB}{0,176,80}
\definecolor{blue}{RGB}{15,158,213}
\definecolor{red}{RGB}{255,0,0}

\usepackage{subfig}

\usepackage{url}

\allowdisplaybreaks 

\begin{document}
\onecolumn{

\large
\noindent Please cite this paper as: \newline
M. Wojnar et al., "Coordinated Spatial Reuse Scheduling With Machine Learning in IEEE 802.11 MAPC Networks," in IEEE Journal on Selected Areas in Communications, doi: 10.1109/JSAC.2025.3584555. 

\normalsize
\vspace{1cm}
\begin{verbatim}

@ARTICLE{wojnar2025coordinated,
  author={Wojnar, Maksymilian and Ciężobka, Wojciech and Tomaszewski, Artur and 
  Chołda, Piotr and Rusek, Krzysztof and Kosek-Szott, Katarzyna and 
  Haxhibeqiri, Jetmir and Hoebeke, Jeroen and Bellalta, Boris and 
  Zubow, Anatolij and Dressler, Falko and Szott, Szymon},
  journal={{IEEE Journal on Selected Areas in Communications}}, 
  title={{Coordinated Spatial Reuse Scheduling With Machine Learning 
  in IEEE 802.11 MAPC Networks}}, 
  year={2025},
  doi={10.1109/JSAC.2025.3584555}
}
\end{verbatim}

}
\twocolumn
\clearpage

\title{Coordinated Spatial Reuse Scheduling With Machine Learning in IEEE 802.11 MAPC Networks}
%
%
%

\author{
Maksymilian Wojnar, 
Wojciech Ciężobka, 
Artur Tomaszewski, Piotr Chołda, 
Krzysztof Rusek,
\\
Katarzyna Kosek-Szott, 
Jetmir Haxhibeqiri\IEEEmembership{~Member,~IEEE}, 
Jeroen Hoebeke, Boris Bellalta\IEEEmembership{~Senior Member,~IEEE},
\\
Anatolij Zubow\IEEEmembership{~Senior Member,~IEEE}, Falko Dressler\IEEEmembership{~Fellow,~IEEE}, 
Szymon Szott
\thanks{M. Wojnar (corresponding author), W. Ciężobka, P. Chołda, K. Rusek, K. Kosek-Szott, and S. Szott are with the Institute of Telecommunications, AGH University of Kraków, Poland, e-mail: \{name.surname\}@agh.edu.pl.}
\thanks{A. Tomaszewski is with the Faculty of Electronics, Telecommunications and Informatics, Gdańsk University of Technology, Poland, e-mail: artur.tomaszewski@pg.edu.pl.}%
\thanks{J. Haxhibeqiri and J. Hoebeke are with the University of Ghent, Belgium, e-mail: \{Jetmir.Haxhibeqiri, Jeroen.Hoebeke\}@UGent.be.}
\thanks{B. Bellalta is with the Department of Engineering, UPF Barcelona, Spain, e-mail: boris.bellalta@upf.edu.}%
\thanks{A. Zubow and F. Dressler are with the School of Electrical Engineering and Computer Science, TU Berlin, 10587 Berlin, Germany, e-mail: \{zubow, dressler\}@tkn.tu-berlin.de.}

}

%
%

\markboth{IEEE Journal on Selected Areas in Communications}%
{IEEE Journal on Selected Areas in Communications}
%



\maketitle

\makeatletter
\def\ps@IEEEtitlepagestyle{%
  \def\@oddfoot{\mycopyrightnotice}%
  \def\@evenfoot{}%
}
\def\mycopyrightnotice{%
  {\footnotesize\hfill 
  {\color{red}
  Accepted for publication in IEEE JSAC. DOI 10.1109/JSAC.2025.3584555 \copyright~IEEE 2025}
  \hfill}%
  \gdef\mycopyrightnotice{}
}
\makeatother

\begin{abstract}
The densification of Wi-Fi deployments means that fully distributed random channel access is no longer sufficient for high and predictable performance. Therefore, the upcoming IEEE 802.11bn amendment introduces multi-access point coordination (MAPC) methods. This paper addresses a variant of MAPC called coordinated spatial reuse (C-SR), where devices transmit simultaneously on the same channel, with the power adjusted to minimize interference. The C-SR scheduling problem is selecting which devices transmit concurrently and with what settings. We provide a theoretical upper bound model, optimized for either throughput or fairness, which finds the best possible transmission schedule using mixed-integer linear programming. Then, a practical, probing-based approach is proposed which uses multi-armed bandits (MABs), a type of reinforcement learning, to solve the C-SR scheduling problem. We validate both classical (flat) MAB and hierarchical MAB (H-MAB) schemes with simulations and in a testbed. Using H-MABs for C-SR improves aggregate throughput over legacy IEEE 802.11 (on average by 80\% in random scenarios), without reducing the number of transmission opportunities per station. Finally, our framework is lightweight and ready for implementation in Wi-Fi devices.
\end{abstract}


\IEEEpeerreviewmaketitle

\section{Introduction}
\IEEEPARstart{F}{or} the past 25 years, channel access in IEEE 802.11 (Wi-Fi) has been based on fully distributed random access.
With network densification, this `no-coordination' approach no longer meets the requirements of recent applications (such as augmented reality).
The first improvement in this area was the introduction of intra-basic service set (BSS) coordination with orthogonal frequency-division multiple access (OFDMA) in 802.11ax (Wi-Fi 6), where the access point (AP) coordinates both uplink (UL) and downlink (DL) transmissions. 
The next step is 802.11bn inter-BSS coordination, where neighboring APs coordinate their transmissions for interference management.

\begin{figure}[!t]
\centering
\subfloat[Topology and AP-station associations.]{
\begin{tikzpicture}[scale=0.5]
            \node[regular polygon, regular polygon sides=3,draw,thick,fill=green,minimum height=0.15cm,minimum width=0.15cm, label=below:{AP\,A}] (ap_1) at (0,0) {};
            \node[rectangle,draw,fill=green,label=below:{Station\,1}] (sta_a) at (-2.5,-0.5) {};
            \draw[dotted,-,line width=1.25] (ap_1) -- (sta_a);
            \node[rectangle,draw,fill=green,label=below:{Station\,2}] (sta_c) at (2.5,-0.5) {};
            \draw[dotted,-,line width=1.25] (ap_1) -- (sta_c);
            \node[regular polygon, regular polygon sides=3,draw,thick,fill=blue,minimum height=0.15cm,minimum width=0.15cm, label=above:{AP\,B}] (ap_2) at (5.5,0) {};
            \node[rectangle,draw,fill=blue,label=above:{Station\,3}] (sta_d) at (3.0,0.5) {};
            \draw[dotted,-,line width=1.25] (ap_2) -- (sta_d);
            \node[rectangle,draw,fill=blue,label=above:{Station\,4}] (sta_e) at (8.0,0.5) {};
            \draw[dotted,-,line width=1.25] (ap_2) -- (sta_e);
\end{tikzpicture}
}
\hfill
\subfloat[Transmission configurations that are viable.]{
\begin{tikzpicture}[scale=0.5]
            \node[regular polygon, regular polygon sides=3,draw,thick,fill=green,minimum height=0.15cm,minimum width=0.15cm] (ap_1) at (0,0) {};
            \node[rectangle,draw,fill=green] (sta_a) at (-2.5,-0.5) {};
            \draw[->,line width=1.5] (ap_1) -- (sta_a);
            \node[rectangle,draw,fill=green] (sta_c) at (2.5,-0.5) {};
            \node[regular polygon, regular polygon sides=3,draw,thick,fill=blue,minimum height=0.15cm,minimum width=0.15cm] (ap_2) at (5.5,0) {};
            \node[rectangle,draw,fill=blue] (sta_d) at (3.0,0.5) {};
            \node[rectangle,draw,fill=blue] (sta_e) at (8.0,0.5) {};
            \draw[->,line width=1.5] (ap_2) -- (sta_e);
            \node[regular polygon, regular polygon sides=3,draw,thick,fill=green,minimum height=0.15cm,minimum width=0.15cm] (ap_1_a) at (0,-2) {};
            \node[rectangle,draw,fill=green] (sta_a_a) at (-2.5,-2.5) {};
            \node[rectangle,draw,fill=green] (sta_c_a) at (2.5,-2.5) {};
            \draw[->,line width=1.5] (ap_1_a) -- (sta_c_a);
            \node[regular polygon, regular polygon sides=3,draw,thick,fill=blue,minimum height=0.15cm,minimum width=0.15cm] (ap_2_a) at (5.5,-2) {};
            \node[rectangle,draw,fill=blue] (sta_d_a) at (3.0,-1.5) {};
            \node[rectangle,draw,fill=blue] (sta_e_a) at (8.0,-1.5) {};
            \node[regular polygon, regular polygon sides=3,draw,thick,fill=green,minimum height=0.15cm,minimum width=0.15cm] (ap_1_b) at (0,-4) {};
            \node[rectangle,draw,fill=green] (sta_a_b) at (-2.5,-4.5) {};
            \node[rectangle,draw,fill=green] (sta_c_b) at (2.5,-4.5) {};
            \node[regular polygon, regular polygon sides=3,draw,thick,fill=blue,minimum height=0.15cm,minimum width=0.15cm] (ap_2_b) at (5.5,-4) {};
            \node[rectangle,draw,fill=blue] (sta_d_b) at (3.0,-3.5) {};
            \draw[->,line width=1.5] (ap_2_b) -- (sta_d_b);
            \node[rectangle,draw,fill=blue] (sta_e_b) at (8.0,-3.5) {};

  \begin{pgfonlayer}{background}
    \filldraw [line width=4mm,join=round,black!10]
      (-2.5,-0.5) rectangle (8.0,0.5)
      (-2.5,-2.5) rectangle (8.0,-1.5)
      (-2.5,-4.5) rectangle (8.0,-3.5);
  \end{pgfonlayer}

  \node at (-2.0,0.5) {\scriptsize\textbf{Case I}};
  \node at (-2.0,-1.5) {\scriptsize\textbf{Case II}};
  \node at (-2.0,-3.5) {\scriptsize\textbf{Case III}};
            
\end{tikzpicture}
}
\hfill
\subfloat[Transmission configurations leading to collisions.]{
\begin{tikzpicture}[scale=0.5]
            \node[regular polygon, regular polygon sides=3,draw,thick,fill=green,minimum height=0.15cm,minimum width=0.15cm] (ap_1) at (0,0) {};
            \node[rectangle,draw,fill=green] (sta_a) at (-2.5,-0.5) {};
            \draw[->,line width=1.5] (ap_1) -- (sta_a);
            \node[rectangle,draw,fill=green] (sta_c) at (2.5,-0.5) {};
            \node[regular polygon, regular polygon sides=3,draw,thick,fill=blue,minimum height=0.15cm,minimum width=0.15cm] (ap_2) at (5.5,0) {};
            \node[rectangle,draw,fill=blue] (sta_d) at (3.0,0.5) {};
            \draw[->,line width=1.5] (ap_2) -- (sta_d);
            \node[rectangle,draw,fill=blue] (sta_e) at (8.0,0.5) {};
            
            \node[regular polygon, regular polygon sides=3,draw,thick,fill=green,minimum height=0.15cm,minimum width=0.15cm] (ap_1_a) at (0,-2) {};
            \node[rectangle,draw,fill=green] (sta_a_a) at (-2.5,-2.5) {};
            \node[rectangle,draw,fill=green] (sta_c_a) at (2.5,-2.5) {};
            \draw[->,line width=1.5] (ap_1_a) -- (sta_c_a);
            \node[regular polygon, regular polygon sides=3,draw,thick,fill=blue,minimum height=0.15cm,minimum width=0.15cm] (ap_2_a) at (5.5,-2) {};
            \node[rectangle,draw,fill=blue] (sta_d_a) at (3.0,-1.5) {};
            \node[rectangle,draw,fill=blue] (sta_e_a) at (8.0,-1.5) {};
            \draw[->,line width=1.5] (ap_2_a) -- (sta_e_a);
            
            \node[regular polygon, regular polygon sides=3,draw,thick,fill=green,minimum height=0.15cm,minimum width=0.15cm] (ap_1_b) at (0,-4) {};
            \node[rectangle,draw,fill=green] (sta_a_b) at (-2.5,-4.5) {};
            \node[rectangle,draw,fill=green] (sta_c_b) at (2.5,-4.5) {};
            \draw[->,line width=1.5] (ap_1_b) -- (sta_c_b);
            \node[regular polygon, regular polygon sides=3,draw,thick,fill=blue,minimum height=0.15cm,minimum width=0.15cm] (ap_2_b) at (5.5,-4) {};
            \node[rectangle,draw,fill=blue] (sta_d_b) at (3.0,-3.5) {};
            \draw[->,line width=1.5] (ap_2_b) -- (sta_d_b);
            \node[rectangle,draw,fill=blue] (sta_e_b) at (8.0,-3.5) {};

  \begin{pgfonlayer}{background}
    \filldraw [line width=4mm,join=round,black!10]
      (-2.5,-0.5) rectangle (8.0,0.5)
      (-2.5,-2.5) rectangle (8.0,-1.5)
      (-2.5,-4.5) rectangle (8.0,-3.5);
  \end{pgfonlayer}

  \node at (-2.0,0.5) {\scriptsize\textbf{Case IV}};
  \node at (-2.0,-1.5) {\scriptsize\textbf{Case V}};
  \node at (-2.0,-3.5) {\scriptsize\textbf{Case VI}};
            
\end{tikzpicture}
}
\caption{Example of the downlink scheduling problem 
in C-SR. For clarity, we omit varying the transmit power. Stations are associated to nearest APs (a). To ensure successful frame reception, APs coordinate to either transmit simultaneously to  outer stations or unilaterally to inner stations (b). Without coordination, if both APs transmit simultaneously and one transmits a frame to an inner station, that frame will collide (c).}
\label{fig:mapc_csr_example}
\end{figure}
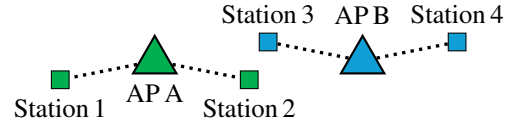
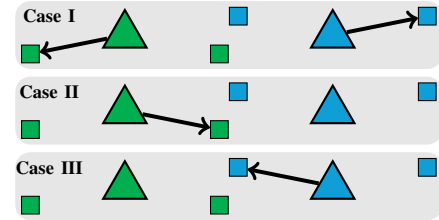
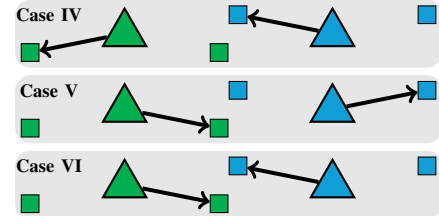

Coordination among neighboring APs (called overlapping BSS, OBSS) improves network performance in terms of higher reliability, reduced latency, increased throughput, and lower power consumption \cite{imputato2024beyond}.
A good example of network settings where such improvements can be observed are factories and shopping malls, where the communication ranges of APs overlap.
In particular, cell edge users, who suffer from interference from neighboring APs, benefit from inter-BSS coordination.
Therefore, a multi-AP coordination (MAPC) feature is planned for 2028 with the release of 802.11bn \cite{giordano2023will}.

The IEEE 802.11bn Task Group lists several modes of MAPC operation in its Specification Framework Document \cite{802.11-24/0209r5}:
coordinated spatial reuse (C-SR),
coordinated beamforming (C-BF),
coordinated time division multiple access, and
coordinated restricted target wake time.
We focus on the first one, formally defined as ``multi-AP Coordinated Spatial Reuse (C-SR) at transmission opportunity (TXOP)-level with power control''. 
This definition means that Wi-Fi devices (APs or stations) transmit simultaneously, on the same channel, with adjusted transmit power to avoid interference.
C-SR is not only pre-approved by the TGbn Task Group, it will also most likely be easy to implement as there are no requirements at the PHY layer.
Theoretically, existing devices would require only driver updates to parse dedicated frames (for over-the-air synchronization), although the exact signaling still remains to be specified.
Note that C-SR is not the same as (regular, uncoordinated) spatial reuse (SR), a feature introduced in 802.11ax.
We explain the differences between SR and C-SR in Section~\ref{sec:csr_operation}.

Consider a C-SR example with two APs and two associated stations per AP (Fig.~\ref{fig:mapc_csr_example}).
In this example and in the remainder of the paper, we focus on DL transmissions (since most traffic is DL, as per the Ericsson Mobility Report of November 2024).
The key feature in this example are the two cell-edge stations~(i.e., stations~2 and~3), which are in the interference range of both APs.
If both APs are coordinated, their transmissions can be successfully decoded at the receiving stations. 
There are three transmission configurations that are viable: simultaneous transmission to the outer stations (case~I) or a unilateral transmission (i.e., with one AP refraining from transmission) to one of the inner stations (cases~II and~III).
Meanwhile, if both APs transmit simultaneously and one of the inner (cell-edge) stations is the intended recipient, that transmission will fail due to interference from the other AP (cases~IV to~VI).
Given that the clear channel assessment (CCA) mechanism on both APs will not detect ongoing transmissions (as they are out of sensing range of each other), without coordination, all six cases may occur indiscriminately.
Even enabling frame protection (i.e., request-to-send/clear-to-send, RTS/CTS, frames) is not a solution because APs will still not transmit simultaneously even if it is viable (e.g., case I in Fig.~\ref{fig:mapc_csr_example} cannot be attained with RTS/CTS).
Thus, the performance of IEEE 802.11 in such topologies is greatly improved with coordinated methods such as C-SR, provided that we determine which transmission configurations are viable.

We refer to the challenge of selecting transmission configurations and transmit power settings in C-SR (i.e., distinguishing cases~I to~III from cases~IV to~VI in Fig.~\ref{fig:mapc_csr_example}), as the \textit{\mbox{C-SR} scheduling problem}.
Our first contribution is a theoretical upper bound model of C-SR (Section~\ref{sec:model}), which finds the best possible transmission schedule using mixed-integer linear programming (MILP).
The model can be optimized for either fairness or throughput and serves as a baseline for comparing practical C-SR scheduling solutions.

C-SR scheduling solutions typically resort to channel measurements (Section~\ref{sec:soa}).
These measurements can be passive, e.g., based on the received signal strength (RSS) of overheard frames, or active, e.g., acquiring channel state information (CSI) from an additional channel sounding frame exchange.
Then, on the basis of these measurements, a decision is made regarding which transmission configurations are viable.
Unfortunately, the aforementioned measurements can be unreliable (RSS), create signaling overhead (both RSS and CSI), require inter-OBSS frame exchange (CSI), and even be unusable and not scalable for C-SR (RSS).
Consider the adjoining research area of data rate selection in Wi-Fi networks, where multiple measurement-based methods have been proposed \cite{yin2020mac}.
These methods have not been adopted due to the previously mentioned issues with RSS, and today wireless drivers typically use probing-based approaches \cite{grunblatt2019simulation}.
Inspired by such developments, we show how to solve the C-SR scheduling problem using a novel probing approach.

We turn to machine learning (ML) and use multi-armed bandit (MAB) methods~\cite{slivkins2019introduction}, to find which transmission configurations are viable in a C-SR setting.
The main benefit of MABs is that they are reinforcement learning (RL) methods which do not require \textit{a priori} knowledge (such as channel measurements) and can learn online, which is suitable for the random and nomadic nature of Wi-Fi deployments. 
We aim to answer the following research questions:
\begin{enumerate}
    \item How quickly can MABs learn which C-SR transmission configurations to use?
    \item How close to optimal are the transmission configurations found?
\end{enumerate}
These research questions aim to address the applicability of using MAB methods in \mbox{Wi-Fi} networks.
We contribute to answering these questions by using two types of MABs: a classical one (flat MAB) and a hierarchical one (H-MAB), both tailored to the C-SR scheduling problem.
Then, we evaluate both MAB C-SR frameworks with extensive simulations in static and dynamic topologies and in an experimental testbed.
For comparison, we use both legacy IEEE 802.11 channel access (the distributed coordination function, DCF) and 802.11ax SR as lower bounds, as well as the provided upper bound model of C-SR.
To encourage research in this area, all of our contributions are available as open-source\footnote{\url{https://github.com/ml4wifi-devs/csr}}. 

We show that the proposed probing-based approach of MABs can be successfully applied to solving the C-SR scheduling problem. 
We show that MAB-based C-SR improves network performance over existing channel access schemes (DCF, SR) achieving high throughput while preserving the minimum number of TXOPs per each station. 
We also validate the proposed framework in an experimental setting, where its adaptability to the underlying network conditions is shown.
Finally, we emphasize that MABs are lightweight and simple to implement and, in contrast to deep RL solutions (which require dedicated processing units), do not have any requirements that prevent them from being implemented in existing devices. 

In summary, our key contributions are as follows:
\begin{enumerate}
    \item An optimization model of C-SR for determining upper bound performance (defined as maximizing either throughput or fairness).
    \item The application of a MAB probing approach to the C-SR scheduling problem coupled with a hierarchical design to tackle scalability issues\footnote{We first applied MABs to C-SR scheduling in \cite{wojnar2025ieee}. Here, we extend the hierarchical MABs with a third level (for transmission power selection).}.
    \item An extensive simulation analysis to study the performance of the proposed approach in comparison to existing baselines.
    \item An experimental testbed study to verify the operation of the proposed approach in real-world devices.    
\end{enumerate}

The remainder of this paper is organized as follows.
We review the state of the art and motivate our approach in Section~\ref{sec:soa}.
Then, we explain in detail how C-SR works, what signaling requirements it has, and how it differs from SR (Section~\ref{sec:csr_operation}).
Next, in Section~\ref{sec:model}, we provide a new optimization model of \mbox{C-SR}, which determines either the upper-bound throughput or the upper-bound fairness. Appendix~\ref{appendix} contains a scalability analysis of this model.
We introduce our MAB C-SR framework in Section~\ref{sec:mabs}, which we then evaluate in simulations (Section~\ref{sec:simulation}) and validate with experiments (Section~\ref{sec:experiments}).
Finally, we discuss our findings and outline future work in Section~\ref{sec:conclusions}.

\section{Related Work}
\label{sec:soa}
The need to coordinate transmissions in wireless networks with multiple base stations is a problem also known in the cellular domain.
Hence, 3GPP proposed coordinated multi-point transmission
(CoMP) in LTE Rel. 11 \cite{lee2012coordinated}.
However, CoMP and MAPC are significantly different: CoMP operates on larger areas with a focus on cell edge users, using a different PHY/MAC stack, and being limited to strictly centralized operation. 
Therefore, we turn our focus to solutions for wireless local area networks.

Several articles on the use of MAPC in Wi-Fi networks have already been published, as surveyed by
\citet{verma2024survey}. 
They provide an overview of MAPC variants (e.g., C-OFDMA, C-SR, C-BF) and comment on future MAPC research directions (e.g., multi-band MAPC, proactive MAPC sounding, blockage-aware coordination, and MAPC scalability).

In terms of solutions to the C-SR scheduling problem, \citet{nunez2022txop} estimate the expected rate of each link and select the top results, making sure that senders and receivers are selected only once.
They later extend their approach into a two-stage algorithm comprising compatible group formation and AP selection (based on various fairness criteria, e.g., most packets to send) \cite{nunez2023multi}. 
Meanwhile, \citet{imputato2024beyond} propose an alternative C-SR scheduling algorithm which 
relies on collecting RSSI information and estimating if the resulting signal to interference plus noise ratio (SINR) of coordinated transmissions allows transmitting data with a given modulation and coding scheme (MCS).
Some works focus only on the power allocation problem in C-SR.
\citet{wang2024research} address this with a federated learning double deep Q-network (FL-DDQN) algorithm. 
APs use a local DDQN for interacting with the environment and then update a global model, considering both throughput and latency requirements.
Meanwhile, \citet{talukder2023enhanced} adjust the transmit power levels by defining an appropriate interference tolerance limit using RSSI levels as input.

While most work on C-SR scheduling is simulation-based, \citet{haxhibeqiri2024a} use a real-world testbed to validate their findings.
Using RSSI reports and historical MCS data, a central controller
chooses C-SR groups for downlink transmissions while adjusting the MCS and transmit power values of APs. 
This research is extended in \cite{haxhibeqiri2024b}, where frame queue organization is additionally considered and AP coverage is divided into circular regions to reduce the number of potential concurrent receivers. 
We use the same testbed in Section~\ref{sec:experiments} (but with our novel, ML-based C-SR scheduling solution).

The existing works mostly use measurements (typically of SINR) as input for the C-SR scheduling problem. 
The most common scenario in which the proposed algorithms are tested is simplistic and assumes static topologies composed of four APs with a low number of associated stations. Typically, state-of-the-art works rely on heuristics and network simulations and focus on downlink transmissions. Furthermore, although Wilhelmi et al. \cite{wilhelmi2023throughput} provide an analytic throughput model based on Markov chains, it is limited to two OBSSs. Thus, we conclude that ML-driven probing-based C-SR scheduling has not yet been tested in complex scenarios, and there are no models which estimate the upper bound C-SR performance under any given settings.

\section{C-SR Operation}
\label{sec:csr_operation}
In this section, we first present the differences between centralized and distributed C-SR and explain why we focus on the former. Then, we describe how transmissions are organized in a C-SR frame exchange and show the impact of using channel measurements. We also describe the differences with respect to SR.
Since C-SR has not yet been standardized and TGbn is currently working on relevant signaling aspects \cite{mapc_setup_scheme}, some details of our description below may change by the final release of the 802.11bn amendment.

\subsection{Initialization and Centralized vs Distributed Approach}
\label{sec:centralized}
A C-SR frame exchange is initialized whenever a C-SR-capable AP wins channel access according to legacy rules (DCF). 
This AP becomes the \textit{sharing AP} and participates in the coordinated transmission in the upcoming TXOP.
The other participants in this TXOP (the \textit{shared APs}) are selected in centralized or distributed way.

In the centralized approach, coordination is done in a central network controller, which has a wired connection to the APs.
This controller obtains and processes monitoring data from the network, decides which transmission configurations are viable, and selects an order of configurations which maximizes performance (we explain the particulars in Section~\ref{sec:model} and the required signaling in Fig.~\ref{fig:h-mabs}).
This centralized approach requires that the APs are under a single administrative domain, which is reasonable in a typical C-SR setting (e.g., a factory, shopping mall, or university campus).

The alternative distributed variant assumes that the 
sharing AP
fully controls channel access for the TXOP and invites shared APs to transmit concurrently.
To execute an informed decision, the sharing AP must exchange data with neighboring APs over a wired link (if the APs are under the same administration domain) or over a wireless link (otherwise).
The former case is conceptually similar to the centralized approach while the latter requires that the APs are within their transmission ranges (e.g., if APs~A and~B in Fig.~\ref{fig:mapc_csr_example} cannot receive each other's control frames and the APs do not share a wired connection, then distributed coordination is impossible).
In the following, we focus on centralized C-SR to provide a proof-of-concept of our proposed MAB C-SR framework and leave the fully distributed case, as the more demanding one, for future work due to the lack of space.

\subsection{Signaling Issues with Channel Measurements}
C-SR scheduling can theoretically be based on RSS or CSI channel measurements.
The former requires two steps: (i) selecting which received frames to measure and (ii) signaling these measurements to the transmitter. 
Consider the example in Fig.~\ref{fig:mapc_csr_example}, assuming APs~A and~B transmit only to the inner stations~2 and~3, respectively. 
RSS measurements need to be performed at each station, but separately for both APs.
This is easy because both APs periodically transmit beacon frames.
However, the next step requires additional UL transmissions to transfer the measurements from the stations to their respective APs.
This signaling overhead, combined with the high variance of RSS \cite{ciezobka2024using}, which can lead to erroneous decisions, both reduce the applicability of such measurements in C-SR.

The CSI case, while providing more reliable measurements, requires even more signaling overhead and additional processing power. 
To obtain CSI, dedicated channel sounding frame exchanges is required over-the-air (not only intra- but also inter-OBSS) to determine the channel quality between each pair of devices.
While researchers investigate channel sounding schemes dedicated to C-SR \cite{verma2024survey}, we abstain from resorting to any channel measurements (based on RSS or CSI) and adopt a new probing approach without additional inter-OBSS over-the-air signaling.

\subsection{Transmission Organization}
We assume a fully coordinated network (i.e., with only \mbox{C-SR} transmissions).
TXOPs are initialized by the sharing AP that wins channel access according to DCF rules.
Then, the network controller instructs other APs about the possibility of joining this coordinated TXOP and, if so, their recipient station and transmit power settings.
C-SR transmissions can start upon reception of this message.
In practice, networks will operate in a mixed mode where not all TXOPs are coordinated, but the goal of this paper is to focus solely on performance during coordinated TXOPs and leave mixed scenarios as future work.

All our assumptions and design choices (DL transmissions, central coordination, no channel measurements) simplify the structure of a C-SR TXOP, which comprises a set of parallel DL transmissions of IEEE 802.11 DATA frames, followed by a set of parallel UL transmissions of IEEE 802.11 acknowledgment (ACK) frames. 
These ACKs are simultaneously transmitted using OFDMA as multi-user block ACKs, which ensures that they do not collide.

\subsection{Comparison with SR}
C-SR can be confused with the SR feature of 802.11ax.
In SR, a device receiving a frame from a neighboring network can disregard it (by applying a more permissive CCA threshold) and consider the channel to be idle, provided that it performs a transmission at a reduced power level to avoid interference with the overlapping transmission.
This leads to two key operational differences: in SR only the neighboring AP reduces its transmit power (while in C-SR any or all APs may reduce this power) and the transmissions are asynchronous.
The outcome of this difference is that SR does not help cell-edge users such as the inner stations in Fig.~\ref{fig:mapc_csr_example} and therefore is not widely used.
C-SR is expected to deliver higher performance gains and is the focus of our study.

\section{Upper Bound Model of C-SR}
\label{sec:model}

To analytically solve the C-SR scheduling problem, i.e., determine the upper-bound transmission schedule under the assumption of a full-buffer traffic model (i.e., each AP always has data to send), we develop a 
MILP optimization model. 
This model is flexible: we show that it can maximize aggregate network throughput (which we call T-Optimal) or fairness (which we call F-Optimal).
Since our goal is to calculate the upper bound performance (in terms of throughput or fairness), as stated before, we do not consider APs operating in mixed-mode (with some TXOPs dedicated to non-C-SR transmissions). 
Furthermore, as in the remainder of the paper, we assume omnidirectional single spatial stream antennas.

Our modeling approach is derived from~\cite{CAPONE2010545,PIORO2014134}.
The authors propose therein a methodology for wireless mesh networks which decouples the design of non-interfering AP--station transmissions from the design of the transmission schedule, i.e., the allocation of time periods during which various sets of non-interfering transmissions (called \textit{transmission sets}) can occur. This methodology incorporates an optimization method known as \textit{column generation} \cite{Spliet2024a}. 
We adapt these concepts to MAPC C-SR as follows.

We model the network using a directed bipartite graph $\mathcal{G}=(\mathcal{V},\mathcal{E})$. The vertex set $\mathcal{V}$ represents nodes and is partitioned into two subsets: $\mathcal{A}$ of APs and $\mathcal{S}$ of stations ($\mathcal{V}=\mathcal{A}\cup\mathcal{S}$ and $\mathcal{A}\cap\mathcal{S}=\emptyset$). The arc set $\mathcal{E}$ denotes potential transmission links between APs and stations ($\mathcal{E}\subseteq\mathcal{A}\times\mathcal{S}$). For each link $e\in\mathcal{E}$, we denote its originating node as $a(e)\in\mathcal{A}$, and the terminating node as $s(e)\in\mathcal{S}$, i.e., $e=(a(e),s(e))$. We also introduce auxiliary sets of node-adjacent links: for each node $v\in\mathcal{V}$, $\delta^+(v)\equiv\{e\in\mathcal{E}:a(e)=v\}$ is the set of links originating at $v$, and $\delta^-(v)\equiv\{e\in\mathcal{E}:s(e)=v\}$ is the set of links terminating at $v$.

Given the complex interplay of interference at receiving stations, our optimization process addresses two critical issues:
\begin{enumerate}
    \item \textit{Determining transmission sets and C-SR settings}: First, select APs that can transmit simultaneously, their recipient stations, as well as transmit power levels and MCS values to achieve successful transmissions under the resulting SINR. 
    Then, return the family $\mathcal{T}$ of transmission sets $t$: $\mathcal{T}=\{t_1,t_2,\dots\}$, where each transmission set $t$ is a vector of throughput values (rates) achieved at the receiving stations.
    \item \textit{Allocating transmission sets}: Given $\mathcal{T}$ and the available time slots, determine the number of slots to assign to each transmission set (i.e., how many slots are designated for transmission sets $t_1$, $t_2$, etc.).
\end{enumerate}

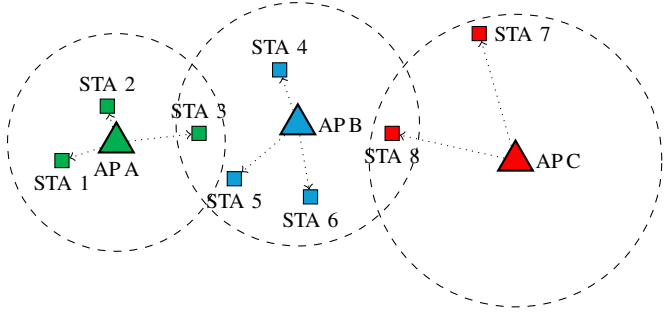
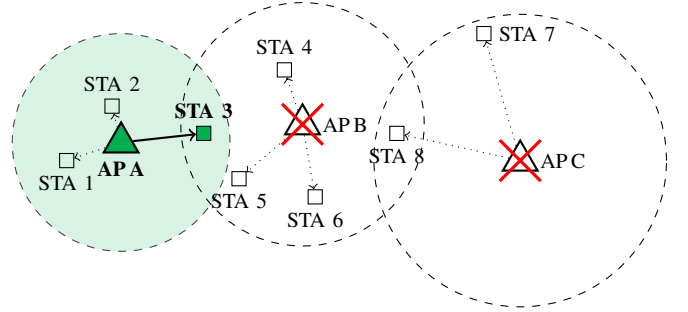
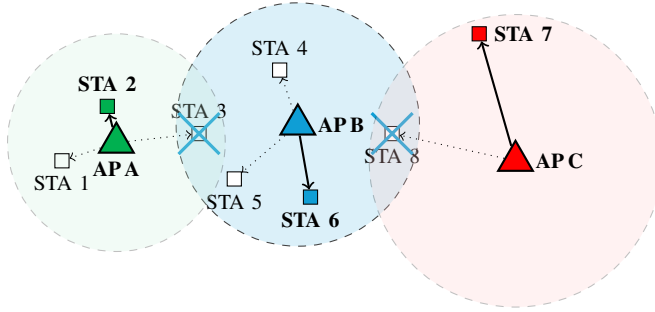
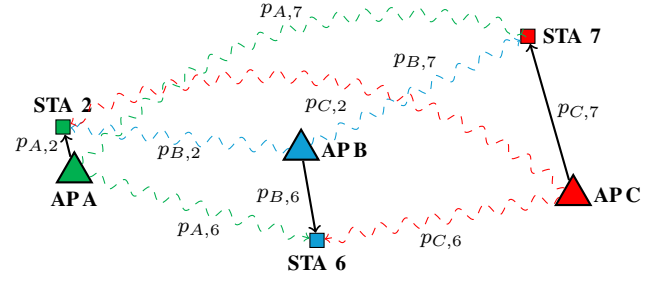
\begin{figure*}
    \footnotesize
    \centering
    \subfloat[APs and their associated stations.]{
        \begin{tikzpicture}[scale=0.48]
            
            \node[regular polygon, regular polygon sides=3,draw,thick,fill=green,minimum height=0.15cm,minimum width=0.15cm, label=below:{AP\,A}] (ap_1) at (0,0) {};
            \coordinate (point1) at (0,3);
            \node[draw, dashed, circle through=(point1)] (b1) at (ap_1) {};
            
            \node[rectangle,draw,fill=green,label=below:{STA~1}] (sta_a) at (-1.5,-0.5) {};
            \draw[dotted,->] (ap_1) -- (sta_a);
            \node[rectangle,draw,fill=green,label=above:{STA~2}] (sta_b) at (-0.25,1) {};
            \draw[dotted,->] (ap_1) -- (sta_b);
            \node[rectangle,draw,fill=green,label=above:{STA~3}] (sta_c) at (2.2775,0.25) {};
            \draw[dotted,->] (ap_1) -- (sta_c);
            
            \node[regular polygon, regular polygon sides=3,draw,thick,fill=blue,minimum height=0.15cm,minimum width=0.15cm, label=right:{AP\,B}] (ap_2) at (5,0.5) {};
            \coordinate (point2) at (2,2);
            \node[draw, dashed, circle through=(point2)] (b2) at (ap_2) {};
            
            \node[rectangle,draw,fill=blue,label=above:{STA~4}] (sta_d) at (4.5,2) {};
            \draw[dotted,->] (ap_2) -- (sta_d);
            \node[rectangle,draw,fill=blue,label=below:{STA~5}] (sta_e) at (3.25,-1) {};
            \draw[dotted,->] (ap_2) -- (sta_e);
            \node[rectangle,draw,fill=blue,label=below:{STA~6}] (sta_e) at (5.35,-1.5) {};
            \draw[dotted,->] (ap_2) -- (sta_e);
            
            \node[regular polygon,
            regular polygon sides=3,draw,thick,fill=red,minimum height=0.15cm,minimum width=0.15cm, label=right:{AP\,C}] (ap_3) at (11,-0.5) {};
            \coordinate (point3) at (7,-1);
            \node[draw, dashed, circle through=(point3)] (b3) at (ap_3) {};
            
            \node[rectangle,draw,fill=red,label=right:{STA~7}] (sta_g) at (10,3) {};
            \draw[dotted,->] (ap_3) -- (sta_g);
            \node[rectangle,draw,fill=red,label=below:{STA~8}] (sta_h) at (7.6,0.25) {};
            \draw[dotted,->] (ap_3) -- (sta_h);
                  
        \end{tikzpicture}
        \label{}
    }
    \hfill
    \subfloat[Exemplary transmission set $t_1$, where only AP\,A transmits and the crossed-out APs are switched off (although due to lack of interference, AP\,C could also transmit).]{
		\begin{tikzpicture}[scale=0.48]
			
			\coordinate (point1) at (0,3);
			\node[draw, fill=green!20, dashed, circle through=(point1),opacity=0.75] (b1) at (0,0) {};
			\node[regular polygon, regular polygon sides=3,draw,thick,fill=green,minimum height=0.15cm,minimum width=0.15cm, label=below:{\bfseries AP\,A}] (ap_1) at (0,0) {};
			
			\node[rectangle,draw,label=below:{STA~1}] (sta_a) at (-1.5,-0.5) {};
			\node[rectangle,draw,label=above:{STA~2}] (sta_b) at (-0.25,1) {};
			\node[rectangle,draw,fill=green,label=above:{\bfseries STA~3}] (sta_c) at (2.2775,0.25) {};
			
			\draw[dotted,->] (ap_1) -- (sta_a);
			\draw[dotted,->] (ap_1) -- (sta_b);
			\draw[thick,->] (ap_1) -- (sta_c);
			
			\node[regular polygon, regular polygon sides=3,draw,thick,fill=white,minimum height=0.15cm,minimum width=0.15cm, label=right:{AP\,B}] (ap_2) at (5,0.5) {};
			\coordinate (point2) at (2,2);
			\node[draw, dashed, circle through=(point2)] (b2) at (ap_2) {};
			
			\node[rectangle,draw,fill=white,label=above:{STA~4}] (sta_d) at (4.5,2) {};
			\draw[dotted,->] (ap_2) -- (sta_d);
			\node[rectangle,draw,fill=white,label=below:{STA~5}] (sta_e) at (3.25,-1) {};
			\draw[dotted,->] (ap_2) -- (sta_e);
			\node[rectangle,draw,fill=white,label=below:{STA~6}] (sta_f) at (5.35,-1.5) {};
			\draw[dotted,->] (ap_2) -- (sta_f);
			
			\node[regular polygon, regular polygon sides=3,draw,thick,fill=white,minimum height=0.15cm,minimum width=0.15cm, label=right:{AP\,C}] (ap_3) at (11,-0.5) {};
			\coordinate (point3) at (7,-1);
			\node[draw, dashed, circle through=(point3)] (b3) at (ap_3) {};
			
			\node[rectangle,draw,fill=white,label=right:{STA~7}] (sta_g) at (10,3) {};
			\draw[dotted,->] (ap_3) -- (sta_g);
			\node[rectangle,draw,fill=white,label=below:{STA~8}] (sta_h) at (7.6,0.25) {};
			\draw[dotted,->] (ap_3) -- (sta_h);
			
			\node[cross out, draw=red, very thick, minimum width=0.5cm, minimum height=0.5cm] at (ap_2) {};

   			\node[cross out, draw=red, very thick, minimum width=0.5cm, minimum height=0.5cm] at (ap_3) {};
			
		\end{tikzpicture}  
        \label{fig:single-transmission-set}
    }
    \\
    \subfloat[Exemplary transmission set $t_2$, where all APs are able to transmit. The crossed-out stations could not be served due to AP\,B transmitting to STA~6.]{
        \begin{tikzpicture}[scale=0.48]
			
			\coordinate (point2) at (2,2);
			\node[draw, fill=blue!20,dashed, circle through=(point2), opacity=0.75] (b2) at (5,0.5) {};
			\node[regular polygon, regular polygon sides=3,draw,thick,fill=blue,minimum height=0.15cm,minimum width=0.15cm, label=right:{\bfseries AP\,B}] (ap_2) at (5,0.5) {};
			\node[rectangle,draw,fill=white,label=above:{STA~4}] (sta_d) at (4.5,2) {};
			\draw[dotted,->] (ap_2) -- (sta_d);
			\node[rectangle,draw,fill=white,label=below:{STA~5}] (sta_e) at (3.25,-1) {};
			\draw[dotted,->] (ap_2) -- (sta_e);
			\node[rectangle,draw,fill=blue,label=below:{\bfseries STA~6}] (sta_f) at (5.35,-1.5) {};
			\draw[thick,->] (ap_2) -- (sta_f);

			\node[rectangle,draw,label=above:{STA~3}] (sta_c) at (2.2775,0.25) {};
			\node[cross out, draw=blue, very thick, minimum width=0.5cm, minimum height=0.5cm] at (sta_c) {};
			\node[rectangle,draw,fill=white,label=below:{STA~8}] (sta_h) at (7.6,0.25) {};
			\node[cross out, draw=blue, very thick, minimum width=0.5cm, minimum height=0.5cm] at (sta_h) {};

			\coordinate (point1) at (0,3);
			\node[draw, fill=green!20, dashed, circle through=(point1),opacity=0.25] (b1) at (0,0) {};
			\node[regular polygon, regular polygon sides=3,draw,thick,fill=green,minimum height=0.15cm,minimum width=0.15cm, label=below:{\bfseries AP\,A}] (ap_1) at (0,0) {};
			
			\node[rectangle,draw,label=below:{STA~1}] (sta_a) at (-1.5,-0.5) {};
			\node[rectangle,draw,fill=green,label=above:{\bfseries STA~2}] (sta_b) at (-0.25,1) {};
			
			\draw[dotted,->] (ap_1) -- (sta_a);
			\draw[thick,->] (ap_1) -- (sta_b);
			\draw[dotted,->] (ap_1) -- (sta_c);

			\coordinate (point3) at (7,-1);
			\node[draw, fill=red!20, dashed, circle through=(point3), opacity=0.25] (b3) at (11,-0.5) {};
			\node[regular polygon, regular polygon sides=3,draw,thick,fill=red,minimum height=0.15cm,minimum width=0.15cm, label=right:{\bfseries AP\,C}] (ap_3) at (11,-0.5) {};
			
			\node[rectangle,draw,fill=red,label=right:{\bfseries STA~7}] (sta_g) at (10,3) {};
			\draw[thick,->] (ap_3) -- (sta_g);
			\draw[dotted,->] (ap_3) -- (sta_h);
			
		\end{tikzpicture}
        \label{fig:triple-transmission-set}
    }
    \hfill
    \subfloat[Possible interference that can decrease the transmission rates. 
    ]{
 	\begin{tikzpicture}[scale=0.6]
			
			\node[regular polygon, regular polygon sides=3,draw,thick,fill=blue,minimum height=0.15cm,minimum width=0.15cm, label=right:{\bfseries AP\,B}] (ap_2) at (5,0.5) {};
			\node[rectangle,draw,fill=blue,label=below:{\bfseries STA~6}] (sta_f) at (5.35,-1.5) {};
			
			\node[regular polygon, regular polygon sides=3,draw,thick,fill=green,minimum height=0.15cm,minimum width=0.15cm, label=below:{\bfseries AP\,A}] (ap_1) at (0,0) {};
			
			\node[rectangle,draw,fill=green,label=above:{\bfseries STA~2}] (sta_b) at (-0.25,1) {};
			
			
			\node[regular polygon, regular polygon sides=3,draw,thick,fill=red,minimum height=0.15cm,minimum width=0.15cm, label=right:{\bfseries AP\,C}] (ap_3) at (11,-0.5) {};
			
			\node[rectangle,draw,fill=red,label=right:{\bfseries STA~7}] (sta_g) at (10,3) {};

			\draw[thick,->] (ap_1) -- node[left] {$p_{A,2}$} (sta_b);
			\draw[thick,->] (ap_2) -- node[left] {$p_{B,6}$} (sta_f);
			\draw[thick,->] (ap_3) -- node[right] {$p_{C,7}$} (sta_g);

			\draw[draw=blue,dashed,->,decorate, decoration=snake] (ap_2) -- node[below] {$p_{B,2}$} (sta_b);
			\draw[draw=blue,dashed,->,decorate, decoration=snake] (ap_2) -- node[above, yshift=0.2cm] {$p_{B,7}$} (sta_g);

			\draw[draw=green,dashed,->,decorate, decoration=snake] (ap_1) -- node[below, yshift=-0.1cm] {$p_{A,6}$} (sta_f);
			\draw[draw=red,dashed,->,decorate, decoration=snake] (ap_3) -- node[below, yshift=-0.1cm] {$p_{C,6}$} (sta_f);

			\draw[draw=green,dashed,->,decorate, decoration=snake] (ap_1) to[bend left] node[above, yshift=0.1cm] {$p_{A,7}$} (sta_g);
			\draw[draw=red,dashed,->,decorate, decoration=snake] (ap_3) to[bend right] node[below, yshift=-0.1cm] {$p_{C,2}$} (sta_b);   
			
		\end{tikzpicture}
        \label{fig:model-interferences}
    }
    \caption{Transmission sets in an exemplary topology. Dashed circles illustrate the potential signal reach for a successful transmission at a given transmit power (the interference range may be larger). The arrows present a potential downlink connection. The vector of throughput obtainable for transmission set $t$ is given as $[r(t,s_1),r(t,s_2),\dots,r(t,s_8)]$, where $r(t,s_i)$ represents the throughput obtained by station~$i$.} \label{fig:transmission-sets}
\end{figure*}

Identifying a transmission set is generally a complex task. While various transmission configurations are theoretically possible, only some are feasible, primarily due to the unacceptable interference from transmissions between different AP--station pairs. 
Since each station is associated with one AP, we identify a link by its receiver $s$. 
Moreover, from an optimization perspective, not all transmission sets are considered advantageous. This concept is illustrated with examples in~Fig.~\ref{fig:transmission-sets}. For example, in Fig.~\ref{fig:single-transmission-set} the exemplary transmission set $t_1$ depends on a single transmission from AP\,A to STA~3. In this scenario, AP\,B cannot transmit to any of its stations due to potential harmful interference with STA~3's reception (STA~3 is in the signal range of AP\,B). Additionally, although AP\,C could transmit, it does not, so that there are no interfering signals for the AP~A~$\rightarrow$~STA~3 transmission. Consequently, the throughput values achieved by stations in relation to this transmission set are represented as $[0,0,r(t_1,s_3),0,0,0,0,0]$, where the bitrate $r(t_1,s_3)$ is determined by SINR at STA~3. Conversely, another transmission set $t_2$ allows three simultaneous transmissions (see Fig.~\ref{fig:triple-transmission-set}), with station downlink throughput values $[0,r(t_2,s_2),0,0,0,r(t_2,s_6),r(t_2,s_7),0]$, where $r(t_2,s_2)$, $r(t_2,s_6)$ and $r(t_2,s_7)$ are derived from the highest feasible MCS achievable during simultaneous transmissions AP\,A~$\rightarrow$~STA~2, AP\,B~$\rightarrow$~STA~6, and AP\,C~$\rightarrow$~STA~7, respectively. However, feasible MCS values should be found based on SINR including interference as shown in Fig.~\ref{fig:model-interferences}, which in some cases (e.g., $p_{C,2}$) may be negligible. We need to find them in the optimization model as well, which makes the task complex.

We break down the overall problem into two sub-problems. The first, \textit{main problem} assumes that transmission sets are predefined (a family of transmission sets is given), focusing on their allocation across different time slots. By reducing the duration of time slots to short intervals, we transition from dealing with discrete time slots to working within a continuous time frame. This transition substantially eases the complexity of the problem. The second, \textit{pricing problem} involves identifying only a single feasible transmission set. If such a new set is found, it extends the family of transmission sets to feed the main problem. The whole procedure is iterated until no new transmission sets are found (Fig.~\ref{fig:loop}). 
A scalability analysis of the model is in Appendix~\ref{appendix}.

Below, we outline how each of these problems is formulated when maximizing \textit{fair} access to the network (F-Optimal), i.e., when we maximize the throughput achieved by the least served station.

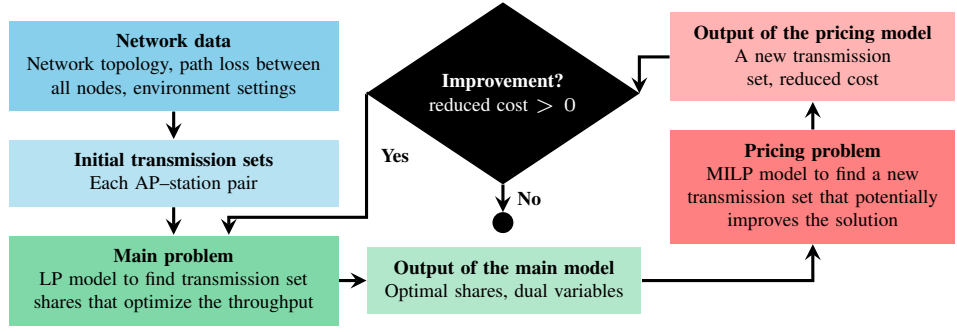
\begin{figure*}
\centering
\resizebox{0.7\textwidth}{!}{%
\begin{tikzpicture}[
    node distance = 0.25cm and 0.25cm,
    every node/.style={text=black,font=\tiny},
    box/.style={rectangle, text width=2.75cm, text centered}, 
    decision/.style={diamond, fill=black, draw, aspect=2, text width=1.35cm, align=center, aspect=1.5},
    arrow/.style={thick,->,>=stealth}
  ]

  \node[box,fill=blue!50] (network_data) {\textbf{Network data } \\ Network topology, path loss between all nodes, environment settings};
  \node[box,fill=blue!30,below=of network_data] (initial_sets) {\textbf{Initial transmission sets} \\ Each AP--station pair};
  \node[box,fill=green!50,below=of initial_sets] (main_problem) {\textbf{Main problem} \\ LP model to find transmission set \\ shares that optimize the throughput};
  \node[box,fill=green!30,right=of main_problem,text width=2.25cm] (output_main) {\textbf{Output of the main model} \\ Optimal shares, dual variables};
  \coordinate (c_help) at ([yshift=-0.40cm]initial_sets.east -| output_main.west);
  \node[box,fill=red!50,right=of output_main,yshift=0.85cm,text width=2.35cm] (pricing_problem) {\textbf{Pricing problem} \\ MILP model to find a new transmission set that potentially improves the solution};
  \node[box,fill=red!30,above=of pricing_problem,text width=2.35cm] (output_pricing) {\textbf{Output of the pricing model} \\ A new transmission set, reduced cost};
  \node[decision, right=of network_data,yshift=-0.25cm, text=white] (improvement) {\textbf{Improvement?} \\ reduced cost $>0$}; 
  \coordinate (c_help1) at ([yshift=0.15cm]improvement.east);
  \node[circle, draw, fill=black, inner sep=0pt, minimum size=5pt, below=of improvement] (end) {};
  
  \draw[arrow] (network_data) -- (initial_sets);
  \draw[arrow] (initial_sets) -- (main_problem);
  \draw[arrow] (main_problem) -- (output_main);
  \draw[arrow] (output_main) -| (pricing_problem);
  \draw[arrow] (pricing_problem) -- (output_pricing);
  \draw[arrow] (output_pricing) -| (c_help1) -- (improvement.east);  
  \draw[arrow] (improvement) -| node[pos=0.75,right,text=black] {\bfseries Yes} (c_help) -|  ([xshift=0.5cm]main_problem.north);
  \draw[arrow] (improvement) -- node[right,text=black] {\bfseries No} (end);

\end{tikzpicture}
}%
\caption{Relations between the main and pricing problems.}
\label{fig:loop}
\end{figure*}

  
\subsubsection{Main Problem}

We operate under the assumption that a normalized time period $[0,1]$ is divided into continuous intervals. In a single interval, exactly one transmission set is executed. For a given family of transmission sets, the objective of the main problem is to identify the optimal mix of time intervals that adheres to a basic fairness criterion. It aims to maximize the achievable received throughput of the station that experiences the lowest performance. 
For each station $s\in\mathcal{S}$, let $R_s$ denote the throughput achieved by station $s$ during the normalized time period, which we seek to optimize through the transmission schedule. Consequently, the problem of transmission scheduling could be formulated as $\max\min_{s\in\mathcal{S}}R_s$. By setting a lower limit on the obtainable throughput $\underline{R}\ge 0$, we can reformulate the latter as a maximization problem: $\max\{\underline{R}:\underline{R}\le R_s, s\in\mathcal{S}\}$, effectively seeking the highest possible minimum throughput that can be applied across all stations.

Let set $\mathcal{T}$ denote the indices of a given family of transmission sets that have been identified so far. Initially, this family might be composed solely of all single-link transmission sets (Fig.~\ref{fig:single-transmission-set}) that can transmit without co-channel interference. 
For each $t\in\mathcal{T}$, let the variable $w_t$ specify the proportion of time (or the duration of a continuous interval) during which the transmission set $t$ is activated. 
Then, assuming that for each $t\in\mathcal{T}$, 
and for every station $s\in\mathcal{S}$, $r(t,s)$ represents the throughput averaged over all slots for station $s$ when transmission set $t$ is activated, the cumulative throughput $R_s$ for station $s$ (accounting for the contributions from the shares of various transmission sets $w_t$) is obtained according to constraints~\eqref{eq:main_problem-throughput_s}. 
Recall that the variable $\underline{R}$ denotes the minimum throughput achievable by each station. Then, we formulate the main problem as the following simple \textit{linear programming} (LP) model:
%
\begin{subequations}\label{eq:main_problem}
\begin{align}
  \max\;        & \underline{R} && \label{eq:main-fairness-optimization}\\
  \text{s.t.}\; & \sum_{t\in\mathcal{T}} w_t = 1 && && |\quad \alpha\\
                & R_s = \sum_{t\in\mathcal{T}} r(t,s) w_t && {s\in\mathcal{S}} && |\quad \beta_s \label{eq:main_problem-throughput_s} \\
                & \underline{R} \le R_s && {s\in\mathcal{S}} \label{eq:main-lower-bound-fairness}
\end{align}
\end{subequations}


From optimization theory, we know that the throughput value of the worst performing station can be potentially improved by expanding the set $\mathcal{T}$ to include a transmission set for which a so-called \textit{reduced cost} 
$\left[ -\alpha+\sum_{s\in\mathcal{S}}r(t,s)\beta_s \right]$ is strictly positive~\cite{Desrosiers2005,Spliet2024a}. Here, $\alpha$ and $\beta_s$, $s\in\mathcal{S}$, represent the optimal values of the dual variables associated with problem~\eqref{eq:main_problem}. These values are accessible once an optimal solution to the main problem has been achieved.

\subsubsection{Pricing Problem}




The \textit{pricing problem} generates (prices out) new variables 
for the restricted main LP problem, improving main's optimum (from the previous iteration). This idea lies at the heart of utilizing the mentioned column generation technique, which facilitates the division of the entire optimization strategy into the two sub-problems. By adopting this method, we can discover an optimal transmission schedule through a cyclic process of solving the main--pricing problem pair. In each cycle, we introduce a new promising transmission set to the family $\mathcal{T}$, continuing this iterative process until the reduced cost obtained from the solution of the pricing problem reaches $0$. The formulation of the pricing problem, shown below, considers  potential interference while various APs transmit. Then, we are able to generate a new transmission set.

For every AP $a\in\mathcal{A}$ and station $s\in\mathcal{S}$, let $l(a,s)$ denote the loss on the transmission path from AP $a$ to station $s$. Let $\mathcal{M}$ be the set of available MCSs. For each mode $m\in\mathcal{M}$, let $r_\mathcal{M}(m)$ and $i(m)$ denote the transmission rate 
and the minimum SINR required by mode $m$, respectively.
Let $\underline{P}$ and $\overline{P}$ denote the minimum and maximum transmit power of an active AP, respectively. $N$ is used to denote the noise floor. 

For each AP $a\in\mathcal{A}$, we define a binary variable $x_a$ that is $1$ if and only if (iff) AP $a$ is selected for transmission. Similarly, for each link $e\in\mathcal{E}$, a binary variable $y_e$ is $1$ iff link $e$ is activated for transmission. Additionally, for every $a\in\mathcal{A}$, a non-negative continuous variable $p_a$ expresses the transmit power of AP $a$. For all links $e\in\mathcal{E}$ and for each mode $m\in\mathcal{M}$, a binary variable $u_{em}$ is $1$ iff transmission mode $m$ is selected for use on link $e$. Furthermore, for each station $s\in\mathcal{S}$, a non-negative continuous variable $r_s$ expresses the downlink throughput of this station. Its optimal value $r_s^*$ will be the throughput for the new transmission set from solving the pricing problem. 

To properly calculate the transmit power, we need to define the SINR condition. To illustrate the related notions, we refer to Fig.~\ref{fig:model-interferences}, and focus on link $e$ representing transmission AP\,B $\rightarrow$ STA~6. Then, we have $a(e) =$ AP\,B, $p_{a(e)} = p_{B,6}$, and the set of other APs: $c\in\mathcal{A}\setminus\{a(e)\} = \{\text{AP\,A},\text{AP\,B}\}$.
We use only the linear relationships related to the signal power levels as well as the SINR requirements. Therefore, we have to use constants $l(x,y)$ related to attenuation levels at $y$ when $x$ is transmitting. 
Then, on the mentioned link $e$, the useful signal level at station $s(e)$, when it receives a signal from AP $a(e)$, is $\frac{p_{a(e)}}{l\left(a(e),s(e)\right)} = \frac{p_{B,6}}{l\left(B,6\right)}$. Continuing our example related to Fig.~\ref{fig:model-interferences}, we should also consider interference from transmissions AP\,A $\rightarrow$ STA~2 and AP\,C $\rightarrow$ STA~7, since in this case $c \in \{\text{AP\,A},\text{AP\,B}\}$. In general, the level of interference and noise is
\begin{equation}
    \zeta_e =\sum_{c\in\mathcal{A}\setminus\{a(e)\}} \frac{p_c}{l\left(c,s(e)\right)}  + N.
\end{equation}
This equation is repeated below in the full problem formulation within constraints~\eqref{eq:zeta}. %
Finally, we state that if the following inequality formulating the SINR condition holds, then we can use mode $m$ in link $e$:
\begin{equation}
    \frac{p_{a(e)}}{l\left(a(e),s(e)\right)} \frac{1}{\zeta_e} \geq i(m). \label{eq:sinr-condition}
\end{equation}

The formulation of the pricing problem, then, is a complex MILP model that incorporates these variables to address the challenge of determining the most efficient transmission configuration under the constraints of the network's physical characteristics and capacities. Simultaneously, this pricing problem considers the dual values from the main problem, as they suggest which transmissions are advantageous. Therefore, this problem is formulated as follows:

\begin{subequations}\label{eq:pricing_problem}
\begin{align}
  \max\;        & - \alpha + \sum_{s\in\mathcal{S}} \beta_s r_s && \label{eq:pricing-goal-fairness}\\
  \text{s.t.}\; & x_a = \sum_{e\in\delta^+(a)} y_e && {a\in\mathcal{A}} \label{eq:pricing_problem-ap_on} \\
                & \sum_{e\in\delta^-(s)} y_e \le 1 && {s\in\mathcal{S}} \label{eq:pricing_problem-station_single} \\
                & p_a \ge \underline{P} x_a && {a\in\mathcal{A}} \label{eq:pricing_problem-station_on_min_energy} \\
                & p_a \le \overline{P} x_a && {a\in\mathcal{A}} \label{eq:pricing_problem-station_on_max_energy} \\
                & \sum_{m\in\mathcal{M}} u_{em} = y_e && {e\in\mathcal{E}} \label{eq:pricing_problem-mcs_link} \\
                & r_s = \sum_{e\in\delta^-(s)}\sum_{m\in\mathcal{M}} r_\mathcal{M}(m) u_{em} && {s\in\mathcal{S}} \label{eq:pricing_problem-station_transfer}\\
                & p_{a(e)} + M(e,m) (1-u_{em}) \ge  \label{eq:pricing_problem-sinr_constraint}\\
                & 
                l\left(a(e),s(e)\right)i(m) \zeta_e && {e\in\mathcal{E}},{m\in\mathcal{M}} \notag  \\
                & \zeta_e =\sum_{c\in\mathcal{A}\setminus\{a(e)\}} \frac{p_c}{l\left(c,s(e)\right)}  + N && {e\in\mathcal{E}} \label{eq:zeta}
\end{align}
\end{subequations}
Constraints~\eqref{eq:pricing_problem-ap_on} make the active AP transmit on exactly one link, and constraints~\eqref{eq:pricing_problem-station_single} forbid the station to receive on more than one link. Constraints~\eqref{eq:pricing_problem-station_on_min_energy} and~\eqref{eq:pricing_problem-station_on_max_energy}  enforce the power of the transmitting AP to be in the range $[\underline{P},\overline{P}]$. Constraints~\eqref{eq:pricing_problem-mcs_link} require that exactly one MCS is used on the transmitting link. Constraints~\eqref{eq:pricing_problem-station_transfer} determine the throughput of the station. 
Finally, constraints~\eqref{eq:pricing_problem-sinr_constraint} are the SINR constrains that must be satisfied if a particular MCS mode is used on the link (and thus $u_{em}=1$). The specific form of constraints~\eqref{eq:pricing_problem-sinr_constraint} improves the numerical properties of the problem by formulating the constraint separately for each MCS. The constraint is deactivated whenever a specific transmission mode $m$ is not used on link $e$ (and thus $u_{em}=0$). In this case, $M(e,m)$ on the left-hand-side makes the inequality trivially true; for all $e\in\mathcal{E}$, $m\in\mathcal{M}$, $M(e,m)$ is a constant equal to the upper bound value of the constraint's right-hand-side expression, i.e., its value when $p_{c}\equiv\overline{P}$, $c\in\mathcal{A}\setminus\{a(e)\}$). Note, that due to constraints~\eqref{eq:pricing_problem-mcs_link} only one of $u_{em}$ for a transmitting link can be 1. Then, $M(e,m)$ is annulled, and the constraint is, in fact, the linearization of the SINR condition given by \eqref{eq:sinr-condition}.

If the optimal objective function value of problem~\eqref{eq:pricing_problem} is strictly positive, the optimal values $r_s^*$ of variables $r_s$, $s\in\mathcal{S}$, define a new transmission set $t'$, which we add to the family $\mathcal{T}$ to resolve the restricted main problem~\eqref{eq:main_problem} (Fig.~\ref{fig:loop}). %

Alternatively, if we decide to optimize for aggregate throughput, then we need to introduce only minor changes. In the main problem, instead of the goal function provided by~\eqref{eq:main-fairness-optimization} we use
$
  \max \; \sum_{s\in\mathcal{S}}{R_s} \label{eq:main-throughput-optimization}
$
and constraints~\eqref{eq:main-lower-bound-fairness} become unnecessary. Additionally, in the pricing problem the goal function given by~\eqref{eq:pricing-goal-fairness} becomes
$
  \max \; - \alpha + \sum_{s\in\mathcal{S}} \left(1+\beta_s\right) r_s. \label{eq:pricing-goal-throughput-optimization}
$

\section{MABs for C-SR}
\label{sec:mabs}

A classical MAB agent learns the reward distribution of each arm by exploring individual choices. 
In the C-SR scheduling problem, each arm is a given transmission configuration (senders, receivers, and transmit power levels) and the reward is the normalized aggregate network throughput.
MAB algorithms implement a trade-off between exploration (knowledge acquisition, i.e., trying new arms) and exploitation (knowledge application, i.e., using arms giving the highest reward).
Since a MAB agent can only make decisions a finite number of times, it should discover the action with the highest reward as quickly as possible.
For these classical MABs, their implementation for C-SR is quite straight forward. 
First, we randomly uniformly select the sharing AP (which models 802.11 channel access, where each AP has a fair chance of winning the contention) and the recipient station (the destination of the head-of-line frame in that AP's transmission buffer).
Then, we enumerate all available choices based on the association between APs and stations, as well as the available transmit power levels, i.e., each configuration is an independent action.
Unfortunately, classical MABs are well suited only for simple topologies (as we will later show) and may fail to find a satisfactory configuration in a reasonable time, since exploration is time-consuming for systems with a large number of arms, as in the case of C-SR in dense topologies (composed of many APs and stations). Therefore, a single action should provide information about the outcomes of similar actions to improve MAB convergence time.
While in the general case this could be difficult, 
we exploit the structure of the underlying C-SR system to construct a hierarchy of actions, which is the key design principle of H-MABs.

To explain this hierarchy, consider a toy example of scheduling pairwise communication between $n$ entities. 
An agent selects who sends to whom and observes the reward of its decision.
The action space consists of (sender, receiver) pairs.
There are $n(n-1)/2$ such pairs and we call this a flat action space, and the agent selecting such pairs -- a flat agent.
Here, the action space is two-dimensional, so a single action is a sequence of two simpler sub-actions: selecting the sender and then the receiver.
This decomposition is the key idea of H-MABs: agents are introduced at each level in the hierarchy where a decision is made.
In our toy example, the first-level agent selects the sender and its action space has $n$ choices. For each sender, we introduce another agent that selects the receiver. This agent has an action space of $n-1$ choices, but there are $n$ such agents, so no action is lost, we only have a different representation.
The key benefit of this new representation is that both agent types have smaller action spaces, while the reward propagates to similar actions. 
In particular, every reward obtained by the downstream agent propagates to the top agent and indirectly affects all actions corresponding to the same sender.
For example, if there is a high reward for allowing particular entities to send, such an agent quickly finds these entities, while the flat agent needs to explore all possible pairs.
Therefore, we design an H-MAB algorithm for C-SR that builds on this idea.

\begin{figure}
\centering
\includegraphics[width=\columnwidth]{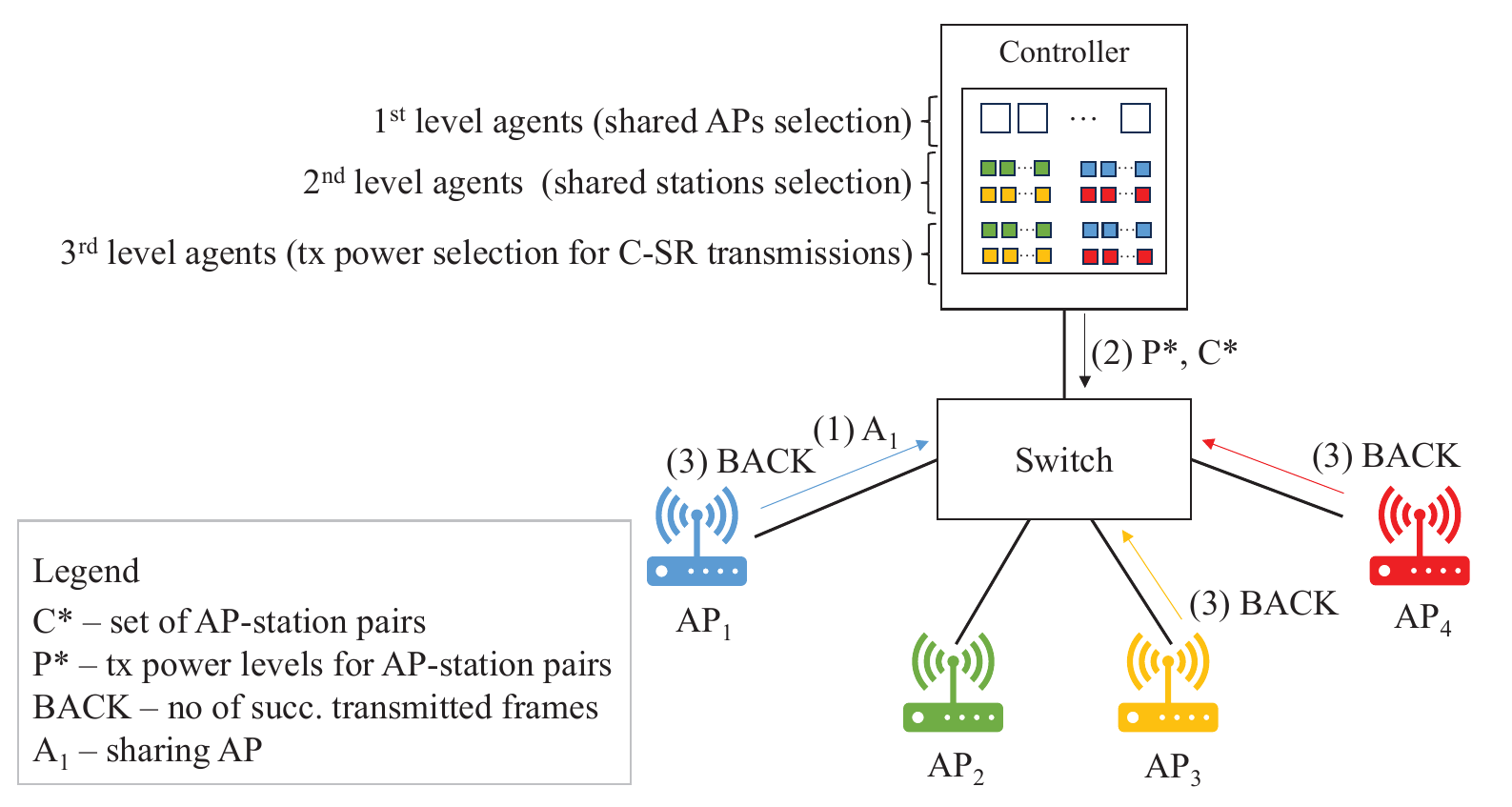}
\caption{Example of C-SR operation with H-MABs: (1) AP~1 notifies the controller that it is the sharing AP, 
(2) the controller provides the set $\mathcal{C}^*$ of AP-station pairs to transmit and the transmit power levels $\mathcal{P}^*$ of the transmitters in the next
TXOP (which include APs~3 and~4), (3) after the TXOP, all transmitting APs
inform the controller how many frames (bytes) were received successfully.}
\label{fig:h-mabs}
\end{figure}

Our H-MAB C-SR framework\footnote{An early version of this framework, with only two levels of agents, appeared in \cite{wojnar2025ieee}.} (Algorithm~\ref{alg:mapc-mab}) considers the network as a set of APs $\mathcal{A}$ with associated stations. 
This assignment partitions the set of all stations into subsets of stations $\mathcal{S}_i = \{S_i^1, S_i^2, \ldots, S_i^m\}$ associated with each AP $A_i$.
In addition, APs can transmit data at any power lever from the set $\mathcal{P}$.
The outcome of the algorithm is a C-SR transmission configuration, comprising the AP--station pairs $\mathcal{C}^*$ and the transmit power levels $\mathcal{P}^*$.
Following the approach described above, we use three steps: selecting the transmitting APs, selecting the recipient stations, and finally selecting the transmit power to use. 
These steps form a three-level hierarchy of agents (Fig.~\ref{fig:h-mabs})\footnote{We formally define our H-MAB framework as an independent, fully cooperative multi-agent RL problem. Although the independent learning framework may not be precise for some problems, such as transmit power selection, which formally requires joint action, it is often used in practice and yields satisfactory results~\cite{Papoudakis2020BenchmarkingMD}. Furthermore, this method significantly decreases the number of actions (from $|\mathcal{P}|^\nu$ to $\nu|\mathcal{P}|$, where $P$ is the set of available transmit power levels and $\nu$ is the number of transmissions), facilitating faster convergence and improving performance in real-world applications.}.
An agent's action is denoted as a sampling operation in Algorithm~\ref{alg:mapc-mab} because the agent is probabilistic and the action is sampled from a distribution.

The algorithm operates on consecutive C-SR TXOPs and begins by randomly uniformly selecting the initial $(A_k, S^l_k)$ pair (lines 2-3). 
Selecting $A_k$ models 802.11 channel access, 
while selecting $S^l_k$ -- the head-of-line frame in that AP's buffer. 
The next decision is whether to add any parallel C-SR transmissions to this pair.

The first-level agent $\alpha^{I}$ (whose parameters depend on the initially selected $(A_k, S^l_k)$ pair) selects the subset of other transmitting APs from all possible subsets $2^{\mathcal{A}\setminus\{A_k\}}$ (line 6).
For all APs that transmit simultaneously $\mathcal{F}$, except the initial one $A_k$, a receiver is selected (line 9).
This is the second-level action and it depends on the previous-level action within the constraints posed by the associations.
Thus, we have the explicit parameters of the second-level agent $\alpha^{II} \left[ A_i, \mathcal{F} \right]$.
Finally, for all C-SR pairs, we select the transmit power level. 
This action is taken by a third-level agent $\alpha^{III} \left[S_i^j, \mathcal{F} \right]$.
The described procedure generates a transmission configuration, which is then executed by performing parallel transmissions according to $\mathcal{C}^*$ with transmit powers set to $\mathcal{P}^*$ and results in an effective data rate $r$ (bytes received across all parallel transmissions during a TXOP).

Since the goal of the proposed H-MAB algorithm for C-SR group selection is to maximize $r$, we use it as the reward and apply a MAB-dependent learning rule denoted as an update.
This update is done in reverse order to the action selection: the third level agent is updated first, then the second level agent, and finally the first level agent.
The reward expressed as the aggregate throughput of all transmissions propagates through the unchanged hierarchy.
This enables the agent, especially its third level, to learn more efficiently by introducing information on the behavior of other agents.
However, the higher the agent's level, the more noise (due to low-level mistakes) in the reward it observes, which is important from the learning algorithm point of view. For this reason, the third level should achieve faster convergence and undertake less exploration. Conversely, the first level should focus on increased exploration and ideally reach convergence subsequent to the third and second levels.
In addition, we assume that MAB agents at each of the three levels are of the same type, although each level has its own configuration (hyperparameter set $\theta$). 

For both flat MABs and H-MABs we tested multiple MAB algorithms ($\epsilon$-greedy, Softmax, Upper Confidence Bound -- UCB, Thompson sampling -- TS) \cite{slivkins2019introduction}, all implemented in Reinforced-lib \cite{wojnar2023reinforced}. We fine-tuned their hyperparameters in random scenarios using Optuna\footnote{\url{https://optuna.readthedocs.io/en/stable/}}. From this comprehensive analysis, we selected the best MAB types for further analysis: Softmax for flat MABs and UCB for H-MABs. 

As explained in Section~\ref{sec:centralized}, we consider a centralized \mbox{C-SR} approach. Fig.~\ref{fig:h-mabs} presents the H-MAB deployment (in case of a flat MAB deployment, only the logic in the controller should be changed). The signaling overhead is small and limited to data transferred over wired links. All computational costs are moved to the controller.
Importantly, H-MABs can also be used in a fully distributed manner, with independent agents deployed at each of the APs (one per AP). 
However, this would require defining additional signaling between APs, which would introduce overhead. Therefore, the applicability of such a distributed solution is left for future work.

\begin{algorithm}[t]
\caption{H-MAB C-SR framework.}
\label{alg:mapc-mab}
\KwIn{\\$\mathcal{A}=\{A_1,A_2,\ldots, A_N\}$ -- set of APs \\
$\mathcal{S}_i = \{S_i^1, S_i^2, \ldots, S_i^m\}$ -- stations associated with $A_i$ \\
$\mathcal{S} = \{\mathcal{S}_1, \mathcal{S}_2, \ldots, \mathcal{S}_N\}$ -- structure of APs and stations \\
$\mathcal{P}=\{P_1,P_2,\ldots, P_W\}$ -- available tx power levels \\
$\theta_1,\theta_2,\theta_3$ -- agents' hyperparameters}
\textbf{Initialize:} \\
$\alpha^I \gets$ the first-level agents based on $\mathcal{A}, \mathcal{S}$ with $\theta_1$ \\ 
$\alpha^{II} \gets$ the second-level agents based on $\mathcal{A}, \mathcal{S}$ with $\theta_2$ \\ 
$\alpha^{III} \gets$ the third-level agents based on $\mathcal{A}, \mathcal{S}, \mathcal{P}$ with $\theta_3$ \\ 
\textbf{Algorithm:} \\
\begin{algorithmic}[1]
\FOR{each TXOP}
\STATE Randomly and uniformly select $A_k \in \mathcal{A}$ 
\STATE Randomly and uniformly select $S_k^l \in \mathcal{S}_k$ 
\STATE $\mathcal{C^*} \gets \{(A_k, S_k^l)\}$ 
\STATE $\mathcal{P^*} \gets \{ \}$ 
\STATE $\mathcal{F}_k^l \in 2^{\mathcal{A}\setminus\{A_k\}} \gets$ sample $\alpha^I \left[ S_k^l \right]$ 
\STATE $\mathcal{F} \gets \mathcal{F}_k^l \cup \{A_k\}$
\FOR{$A_i \in \mathcal{F}_k^l$}
\STATE $S_i^j \in \mathcal{S}_i \gets$ sample $\alpha^{II} \left[ A_i, \mathcal{F} \right]$
\STATE $\mathcal{C^*} \gets \mathcal{C^*} \cup \{(A_i, S_i^j)\}$
\ENDFOR
\FOR{$A_i, S_i^j \in \mathcal{C^*}$}
\STATE $P^j \in \mathcal{P} \gets$ sample $\alpha^{III} \left[S_i^j, \mathcal{F} \right]$
\STATE $\mathcal{P^*} \gets \mathcal{P^*} \cup \{\mathcal{P}^j\}$
\ENDFOR
\STATE Perform simultaneous transmissions according  to $\mathcal{C^*}$ with transmit power levels $\mathcal{P^*}$.
\STATE $r \gets$ effective data rate in the TXOP
\FOR{$A_i, S_i^j \in \mathcal{C^*}$}
\STATE Update agent $\alpha^{III} \left[S_i^j, \mathcal{F} \right]$ with $r$
\IF{$A_i \neq A_k$}
\STATE Update agent $\alpha^{II} \left[A_i, \mathcal{F} \right]$ with $r$
\ENDIF
\ENDFOR
\STATE Update agent $\alpha^I \left[ S_k^l \right]$ with $r$
\ENDFOR
\end{algorithmic}
\label{alg1}
\end{algorithm}

\section{Performance Evaluation} 
\label{sec:simulation}


We compare the performance of our proposed MAB C-SR schemes (flat MAB, H-MAB) with the following baselines:
\begin{itemize}
    \item \textbf{DCF} -- the basic IEEE 802.11 channel access mechanism,
    \item \textbf{SR} -- the OBSS Power Detect SR variant of 802.11ax,
    \item \textbf{F-Optimal} and \textbf{T-Optimal} -- the upper bound model (Section~\ref{sec:model}) optimized for either fairness or throughput.
\end{itemize}
We use our custom-built open source simulators, with all design details provided online\footnote{\url{https://github.com/ml4wifi-devs/csr}}.

\begin{figure} 
  \centering
  \subfloat[Multi-room topology\label{fig:indoor_topology}]{\includegraphics[width=0.5\linewidth]{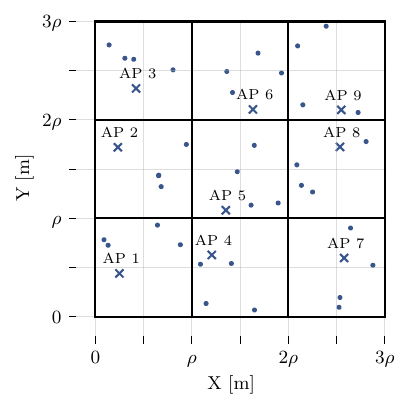}}
  \hfill
  \subfloat[Open space topology\label{fig:free_space_topology}]{\includegraphics[width=0.5\linewidth]{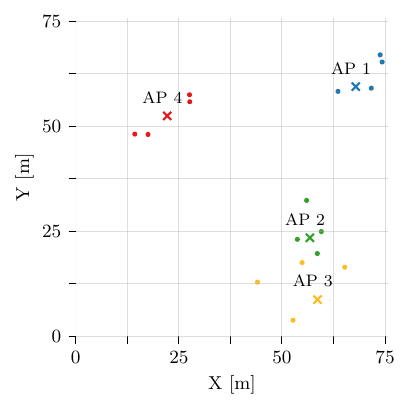}}
  \caption{Exemplary topologies: (a) multi-room and (b) open space. Crosses denote APs, dots -- stations, thick lines -- walls. In (a), an AP and four stations are all randomly placed in each of the $\rho \times \rho$~m rooms. In (b), the APs are placed randomly, while the stations are placed randomly near their APs.}
  \label{fig:scenario_residential}
\end{figure}

We focus on two types of overlapping topologies: multi-room and open space (Fig.~\ref{fig:scenario_residential}). 
The \emph{multi-room} topologies 
have a varying floor size (from 2x2 up to 4x4 square rooms) and varying room sizes (separated by walls). 
Each room contains a single AP and four stations, all uniformly distributed. 
In the \emph{open space} topologies, the positions of the APs are drawn randomly within a square $\SI{75}{\meter} \times \SI{75}{\meter}$ area without walls. Additionally, station positions are randomly selected from a normal distribution with a mean equal to the AP's position and a variable standard deviation. In each open-space tested topology there are 2--5 APs and each AP has 3--5 associated stations. 
Further details on all scenarios are available in the aforementioned repository.


The simulation parameter settings are given in Table~\ref{tab_simulation_settings}. 
We assume, both in these simulations and in our testbed (Section~\ref{sec:experiments}), that all APs operate on the same \SI{20}{\mega\hertz} channel (channel allocation being an orthogonal problem) and use single-stream transmissions. Wider channels and multiple spatial streams would impact the presented results only quantitatively.
For data rate selection, we match the MCS used with the expected SINR at the receiver. 
Each result presented is the average of 10 independent runs. The bands around the lines and the data points denote the 95\% confidence intervals of the mean.


\begin{figure*}[!t]
\centering
\includegraphics[width=\textwidth]{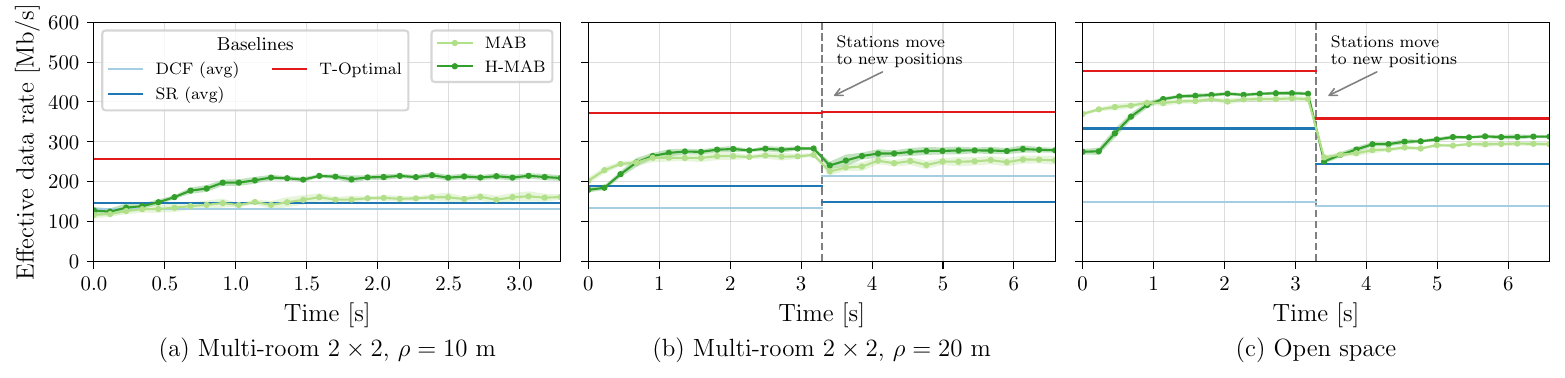}
\caption{Aggregate effective data rate for two multi-room and one  open space topology. 
}
\label{fig:small_office_results}
\end{figure*}

\begin{table}
\caption{Simulation parameter settings}
\label{tab_simulation_settings}
\centering
\begin{tabular}{@{}ll@{}}
\toprule
\textbf{Parameter}         & \textbf{Value}            \\ \midrule
Band                       & \SI{5}{\giga\hertz}                   \\
PHY                    & IEEE   802.11ax           \\
Channel   width            & \SI{20}{\mega\hertz}                  \\
Spatial   streams          & 1, SISO                   \\
Guard   interval           & \SI{800}{\nano\second}                  \\
Frame   aggregation        & Enabled                   \\
TXOP duration         & \SI{5.484}{\milli\second} \\
Path loss model               & TGax enterprise \cite{pathloss} \\
Multipath fading & Additive white Gaussian noise \\
Maximum transmit power         & \SI{16}{dBm}         \\
Noise floor                & \SI{-94}{dBm}          \\
OBSS\_PD threshold  & \SI{-72}{dBm} \\
Wall penetration loss & \SI{7}{dB} \\
Per-station   traffic load & Downlink, full buffer     \\
Frame size & \SI{1500}{B}                  \\ 
Independent repetitions     & 10                        \\ \bottomrule
\end{tabular}
\end{table}

Furthermore, we introduce a heuristic to detect the convergence of MAB agents based on Holt's exponential smoothing \cite{HOLT20045}. We fit the model to the mean value of all runs, then extract the trend, normalize, and smooth. When we detect that the trend is lower than a scenario-dependent threshold for a certain number of steps (i.e., that the throughput achieved by the MAB-driven network has stabilized), we consider this moment as the convergence point. 

\subsection{Transient Performance}
We first compare the transient performance of MABs and H-MABs with three baselines (DCF, SR, and T-Optimal).
In Fig.~\ref{fig:small_office_results}a, when all stations are static, both our approaches converge equally fast although H-MABs ultimately achieve a higher effective data rate.
In Fig.~\ref{fig:small_office_results}b, we introduce nomadic mobility (all nodes change position within their rooms halfway through the simulation), but both MAB architectures quickly adjust to the new conditions. Additionally, in both cases, they outperform DCF and SR and approach the T-Optimal value. 
In the open space topology with a random position change (Fig.~\ref{fig:small_office_results}c), the convergence time for H-MABs is higher than for flat MABs, although their overall performance is similar to each other.
These results also show that MABs and H-MABs can learn a new best C-SR scheduling configuration without resetting their state.
Again, both outperform DCF and SR while remaining close to T-Optimal.
As we show later, the observed similar behavior of (flat) MABs and H-MABs significantly depends on the underlying topology.

We see (both here and in later results) that SR does not always outperform DCF.
The benefits of SR appear only when the separation between APs is large enough to enable SR-based simultaneous transmissions. Otherwise, decreased performance is a known effect \cite{natkaniec2023analysis}. 

\begin{figure}[!t]
\centering
\includegraphics[width=0.95\columnwidth]{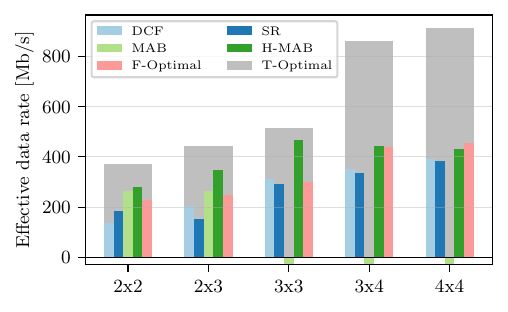}
\caption{Performance in multi-room topologies with $\rho=\SI{20}{\meter}$. Negative bars indicate unrealistic computational requirements. Confidence intervals are not shown, as they are too small to be visually distinguishable.}
\label{fig:results_residential_d20}
\end{figure}

\begin{table}[]
\caption{Average steps needed to converge in multi-room topologies with $\rho=\SI{20}{\meter}$ and their corresponding convergence times for short (\SI{0.1}{\milli\second}), medium (\SI{1}{\milli\second}), and long (\SI{5.484}{\milli\second}) TXOP durations. \emph{inf} indicates that no advantageous configurations where found in a limit of \SI{200000}{} steps.}
\label{tab_convergence}
\centering
\begin{tabular}{@{}lllllll@{}}
\toprule
                                                                 &       & \multicolumn{5}{c}{\textbf{Scenario size}}                               \\ 
                                                                 &       & \textbf{2x2} & \textbf{2x3} & \textbf{3x3} & \textbf{3x4} & \textbf{4x4} \\
                                                                 \midrule
\multirow{2}{*}{\textbf{Steps}}                                  & MAB   & \SI{540}{}          & \SI{1320}{}         & N/A          & N/A          & N/A          \\
                                                                 & H-MAB & \SI{690}{}          & \SI{1680}{}         & \SI{14400}{}        & \SI{120000}{}       & inf          \\\midrule
\multirow{2}{*}{\textbf{\begin{tabular}[c]{@{}l@{}}Short\\ TXOP\end{tabular}} }  
 & MAB   & \SI{0.05}{\second}       & \SI{0.13}{\second}       & N/A          & N/A          & N/A          \\
                                                        & H-MAB & \SI{0.07}{\second}       & \SI{0.17}{\second}       & \SI{1.44}{\second}       & \SI{12.00}{\second}      & inf          \\\midrule
\multirow{2}{*}{\textbf{\begin{tabular}[c]{@{}l@{}}Medium\\ TXOP\end{tabular}}}& MAB   & \SI{0.54}{\second}       & \SI{1.32}{\second}       & N/A          & N/A          & N/A          \\
                                                        & H-MAB & \SI{0.69}{\second}       & \SI{1.68}{\second}       & \SI{14.40}{\second}      & \SI{120.00}{\second}     & inf          \\\midrule
\multirow{2}{*}{\textbf{\begin{tabular}[c]{@{}l@{}}Long\\ TXOP\end{tabular}}}  & MAB   & \SI{2.96 }{\second}      & \SI{7.24 }{\second}      & N/A          & N/A          & N/A          \\
                                                        & H-MAB & \SI{3.78}{\second}       & \SI{9.21}{\second}       & \SI{78.97}{\second}      & \SI{658.08}{\second}     & inf          \\ \bottomrule
\end{tabular}
\end{table}



\begin{figure}[!t]
\centering
\includegraphics[width=0.95\linewidth]{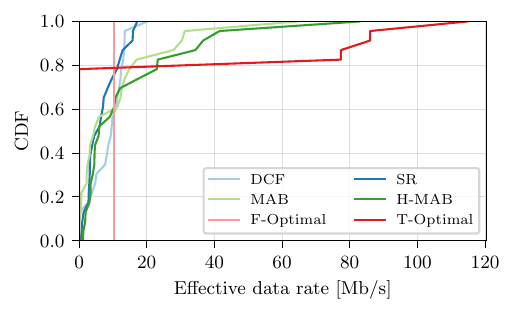}
\caption{CDFs for the 2x3 multi-room topology with $\rho = \SI{20}{\meter}$. The effective data rate for each station is the average of 10 independent runs.}
\label{fig:residential_cdf_2_3_4_d20}
\end{figure}

\subsection{Steady-State Performance in Multi-Room Topologies}
\label{sec:steady_state}
Next, we focus on steady-state performance.
For the multi-room topology, we analyze varying floor configurations (different number of rooms, different room sizes). 
We present results for $\rho=20$~m (Fig.~\ref{fig:results_residential_d20}); results for other $\rho$ values are qualitatively similar and available in our online repository. 
\mbox{H-MABs} outperform the baselines (DCF, SR) as well as flat MABs in most of the room configurations tested. 
T-Optimal achieves better rates than H-MABs since it does not account for any minimum fairness guaranteed by our approach (lines 2-3 of Algorithm~\ref{alg:mapc-mab}), while F-Optimal achieves worse rates because it prioritizes fairness.
This is visible in Fig.~\ref{fig:residential_cdf_2_3_4_d20}, which presents a cumulative distribution function (CDF) of the stations' effective data rates in a 2x3 multi-room configuration for $\rho=\SI{20}{\meter}$. 
Importantly, in all cases shown in Fig.~\ref{fig:results_residential_d20}, H-MABs outperform DCF and SR. The lower gain of C-SR compared to DCF and SR in the 3x4 and 4x4 topologies results from the suboptimal performance of H-MABs for such complex scenarios. H-MABs would require more time to converge to a better solution.

In fact, the convergence times of MABs and H-MABs  (Fig.~\ref{fig:results_residential_d20}) increase with topology size and TXOP duration (Table~\ref{tab_convergence}). Flat MABs have too demanding computational requirements if the number of APs exceeds six due to the extremely large action space, which is at least partially solved by the proposed H-MABs (up to 12 APs with four associated stations each). 
These results clearly show that in large and highly congested deployments, another level of C-SR operation is required, i.e., the creation of independent C-SR clusters managed by separate MAB/H-MAB agents, as we show in Section~\ref{sec:clustered-mabs}. 
This clustering procedure can reduce the size of the action space for a particular sharing AP (e.g., the number of considered shared APs) as well as the convergence time. 
Additionally, recent 802.11bn advancements \cite{802.11-25/0502r0} suggest that C-SR agreements can be created for long-term operation, and then the convergence time is less critical considering the large possible gains (e.g., the ones observed for the 2x2, 2x3, and 3x3 topologies). 

\begin{figure}[!t]
\centering
\includegraphics[width=0.95\columnwidth]{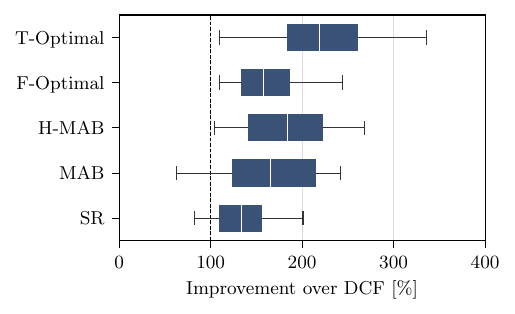}
\caption{Effective data rate improvement averaged over 24 random open space topologies with a random topology change halfway through each simulation.}
\label{fig:random}
\end{figure}

\subsection{Performance in Random Open Space Environments}
We next compare different channel access methods in open space topologies in terms of the average effective data rate (Fig.~\ref{fig:random}). We introduce a random topology change halfway through each simulation, where all nodes are displaced to new positions. This sudden change forces C-SR schemes to adapt to completely different network topologies. These results confirm that (i) both MAB-based approaches outperform not only DCF (on average by 80\%) but also legacy SR, (ii) only H-MABs consistently outperform DCF (there are no outliers below the 100\% line), and (iii) H-MABs are on average better than flat MABs. Additionally, T-Optimal and F-Optimal give results similar to those for the multi-room topologies.

\subsection{Clusterization for Larger Topologies}
\label{sec:clustered-mabs}

\begin{figure} 
  \centering
  \subfloat[3x4 topology\label{fig:clustered_3x4}]{\includegraphics[width=\linewidth]{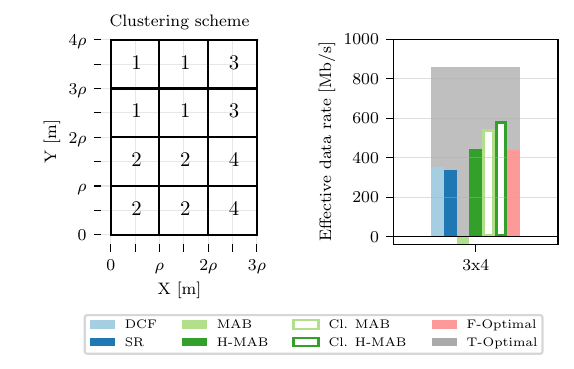}}
  \\
  \subfloat[4x4 topology\label{fig:clustered_4x4}]{\includegraphics[width=\linewidth]{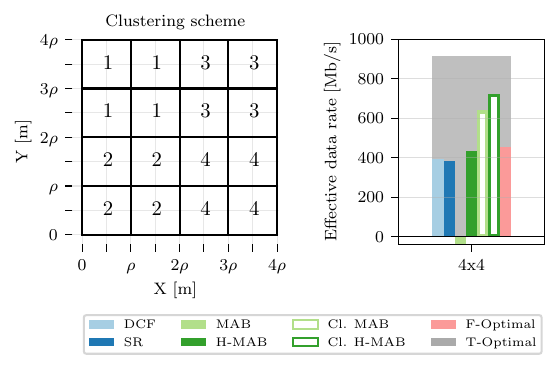}}
  \caption{Performance of clustered MAB and H-MAB variants in the (a)~3x3 and (b) 4x4 multi-room topologies (for $\rho=20$~m). The numbers in each room identify the assigned cluster.}
  \label{fig:clusterization}
\end{figure}

For the highly congested multi-room topologies (3x4, 4x4) analyzed in Section~\ref{sec:steady_state} for which the convergence times (Table~\ref{tab_convergence}) were particularly large, we implement a C-SR clusterization scheme (shown in the left part of Fig.~\ref{fig:clusterization}). 
Each cluster is governed by a single independent MAB or H-MAB agent, so this remains a centralized approach. 
Recall that in all the cases analyzed, the APs share the same wireless channel. 
Therefore, this setting shows the operation of clustered C-SR under strenuous conditions (i.e., the contention level does not change compared to the cases analyzed in Section~\ref{sec:steady_state}).

The effective data rate values presented in Fig.~\ref{fig:clusterization} are much higher for clustered C-SR, not only in comparison to the baselines (DCF and SR) but also in comparison to the clusterless C-SR schemes which were unable to find a satisfactory solution in a reasonable time. Both effective data rates improve and convergence times drastically decrease (Table \ref{tab_convergence-clustered}). The changes are especially visible for the (flat) MAB-based C-SR. Without clusterization, it was unable to converge as a result of the extremely large action space. With the clustered approach, it converges quickly and results in data rates comparable to those of H-MABs.

\begin{table}[]
\caption{Average steps needed by the clustered C-SR to converge in multi-room topologies with $\rho=\SI{20}{\meter}$ and their corresponding convergence times for short (\SI{0.1}{\milli\second}), medium (\SI{1}{\milli\second}), and long (\SI{5.484}{\milli\second}) TXOP durations.}
\label{tab_convergence-clustered}
\centering
\begin{tabular}{@{}llllll@{}}
\toprule
&       & \multicolumn{2}{c}{\textbf{Scenario size}}                               \\             &       & \textbf{3x4} & \textbf{4x4} \\
                                \midrule
\multirow{2}{*}{\textbf{Steps}}                     & MAB   &  \SI{1080}{}          & \SI{760}{}    \\
     & H-MAB & \SI{960}{}          & \SI{1040}{}    \\
                                \midrule
\multirow{2}{*}{\textbf{\begin{tabular}[c]{@{}l@{}}Short\\ TXOP\end{tabular}} }               & MAB   &  \SI{0.11}{\second}        & \SI{0.08}{\second}  \\
 & H-MAB & \SI{0.10}{\second}        &  \SI{0.10}{\second} \\
                                \midrule
\multirow{2}{*}{\textbf{\begin{tabular}[c]{@{}l@{}}Medium\\ TXOP\end{tabular}} }             & MAB   &  \SI{1.08}{\second}        & \SI{0.76}{\second}  \\
     & H-MAB & \SI{0.98}{\second}        & \SI{1.04}{\second}  \\
                                \midrule
\multirow{2}{*}{\textbf{\begin{tabular}[c]{@{}l@{}}Large\\ TXOP\end{tabular}} }              & MAB   &  \SI{5.92}{\second}        & \SI{4.16}{\second}  \\
& H-MAB & \SI{5.26}{\second}     & \SI{5.70}{\second}  \\
\bottomrule
\end{tabular}
\end{table}

\begin{figure}
    \centering
    \includegraphics[width=0.5\linewidth]{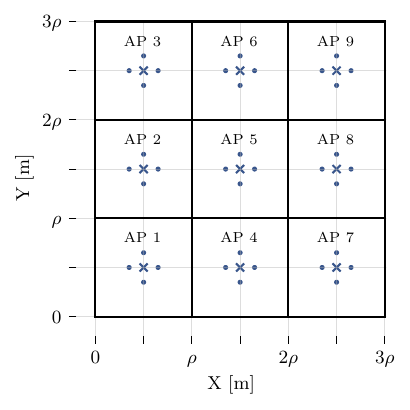}
    \caption{Exemplary symmetrical topology. Crosses denote APs, dots -- stations, thick lines -- walls. Each BSS is composed of an AP and four stations are symmetrically placed at a distance of \SI{2}{\meter}. }
    \label{fig:clustered-symmetrical}
\end{figure}

\begin{figure}[t]
    \centering
    \includegraphics[width=0.95\linewidth]{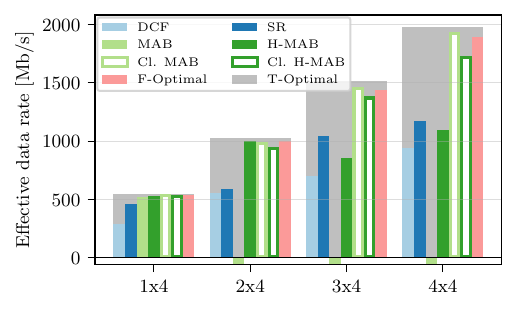}
    \caption{Performance in the symmetrical enterprise scenario. Confidence intervals are not shown, as they are too small to be visually distinguishable.}
    \label{fig:clustered-symmetrical-results}
\end{figure}

Next, we move to more favorable, symmetrical 
large topologies, inspired by the IEEE 802.11ax enterprise scenario \cite{pathloss}, where each BSS consists of an AP with full buffers and four associated stations, placed at a distance of \SI{2}{\meter} (Fig.~\ref{fig:clustered-symmetrical}). 
Neighboring APs are placed at a distance of $\rho=30$~m from each other. 
This topology is favorable for C-SR, because it allows parallel transmissions for all APs during the same TXOP (if proper transmission power levels are configured).

For each topology, we implement a clustered C-SR scheme as follows. For the 3x4 and 4x4 cases, the clusterization is the same as in the 3x4 and 4x4 multi-room topologies (Fig.~\ref{fig:clusterization}). Furthermore, in the 1x4 and 2x4 cases, two clusters are created (each composed of two and four APs, respectively). Each cluster is governed by an independent MAB/H-MAB agent. In such a regular topology, the performance of DCF is the worst (because of its strict CCA threshold) and is only partially improved by SR, which relaxes this constraint (Fig.~\ref{fig:clustered-symmetrical-results}). However, the performance of clustered C-SR with MAB and H-MAB independent agents is much better in terms of the effective data rate, which comes close to that of F-Optimal and T-Optimal (which are similar on account of the favorable conditions mentioned before). Importantly, the performance of both clusterless MAB and H-MAB C-SR is worse than that of SR, because of the significant time that they require to converge to the optimum.

A full implementation of a clustered C-SR approach (i.e., with dynamic cluster creation) requires modifications in the presented C-SR scheme (at the network controller level) and is left for future study. However, this subsection shows the potential of clustered C-SR operation and signalizes an open research challenge.

\subsection{Impact of Legacy Devices}

\begin{figure}
    \centering
    \includegraphics[width=0.95\columnwidth]{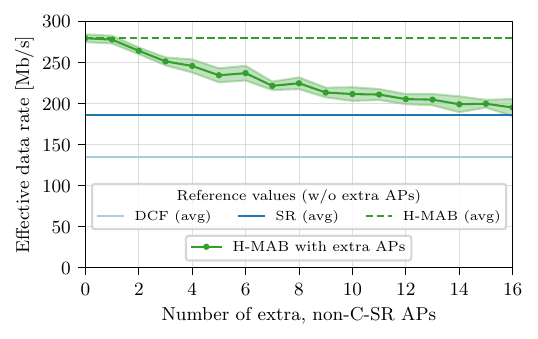}
    \caption{Impact of additional legacy APs on C-SR operation compared to baseline DCF and SR algorithms operation in less crowded environment (i.e., without additional legacy APs) in the 2x2 multi-room topology with $\rho=\SI{20}{\meter}$.}
    \label{fig:legacy}
\end{figure}

So far, we have analyzed the performance of the proposed C-SR scheme under no external interference (i.e., an isolated case). Here, we measure the impact of additional legacy APs on C-SR operation in an exemplary 2x2 multi-room topology (Fig.~\ref{fig:scenario_residential}a). We proceed to uniformly add legacy (non-C-SR) APs so that each of these extra APs contends with exactly one of the four C-SR APs. For example, if there are seven non-C-SR APs, then three of the C-SR APs contend with two extra APs, while the fourth one contends with one extra AP.
As reference levels, we use the performance of DCF and SR, both operating in a network topology without the extra APs. Therefore, in Fig.~\ref{fig:legacy}, the presented performance of DCF and SR as well as the data point for H-MAB and zero legacy APs, corresponds to the results of the 2x2 room in Fig.~\ref{fig:results_residential_d20}. 
We analyze the effective data rate achieved under the condition that one of the C-SR APs successfully wins the channel contention. 
These results show that when the number of legacy APs increases, the operation of C-SR steadily deteriorates due to an increased probability of collisions and fewer opportunities for successful C-SR transmissions. However, even with multiple additional APs the performance of C-SR is better than that of SR and DCF working in a less crowded environment. This suggests that C-SR can bring important gains and improve overall channel utilization, not only for homogeneous, but also heterogeneous deployments. 

\section{Experimental Validation}
\label{sec:experiments}
To validate our simulation findings, we evaluate our proposed MABs in experiments. In the following, we describe the testbed composition and the results obtained.

\subsection{Testbed Description}
We use the Industrial IoT Lab\footnote{\url{https://idlab.ugent.be/resources/industrial-iot-lab}} from Ghent University to perform a real-world validation of our approach. All Wi-Fi devices use openwifi\footnote{\url{https://github.com/open-sdr/openwifi}}, which is the first open-source implementation of Wi-Fi based on software defined radio (SDR). The design includes an open-source implementation of the physical layer, the lower and higher MAC layers, and is fully compatible with off-the-shelf Wi-Fi nodes. 

In addition to the normal Wi-Fi features and flexible parameter configuration, openwifi supports accurate time synchronization with microsecond accuracy \cite{jiao2020openwifi} as well as a traffic scheduling mechanism that assigns dedicated time slots to hardware queues on a periodic basis \cite{haxhibeqiri2021bringing}. In addition, openwifi allows selecting MCS and transmit power values on a per-time slot basis and remotely configuring the SDR devices using a dedicated controller. The duration of these time slots is fully configurable. 
With these main enablers, the openwifi-based testbed is suitable to perform tests of concurrent downlink transmissions from multiple APs (C-SR) \cite{haxhibeqiri2024a} and, therefore, is an ideal candidate to test our proposed solution in practice. 
We optimize MAB and H-MAB hyperparameters in a digital-twin of the testbed recreated in our simulation environment.
Then, we install (flat) MABs and H-MABs on the network controller, which communicates the decisions on the transmission configurations and transmit power settings to APs using Ethernet connections, as shown in Fig.~\ref{fig:h-mabs}. 
The controller also collects information whether the performed transmissions were successful or not, which then serves as a reward for the implemented agents. 
Additionally, a dedicated switch is used to synchronize all of the APs in the network, while the stations are synchronized over the air. 

\begin{figure}
\centering
\includegraphics[width=0.95\columnwidth]{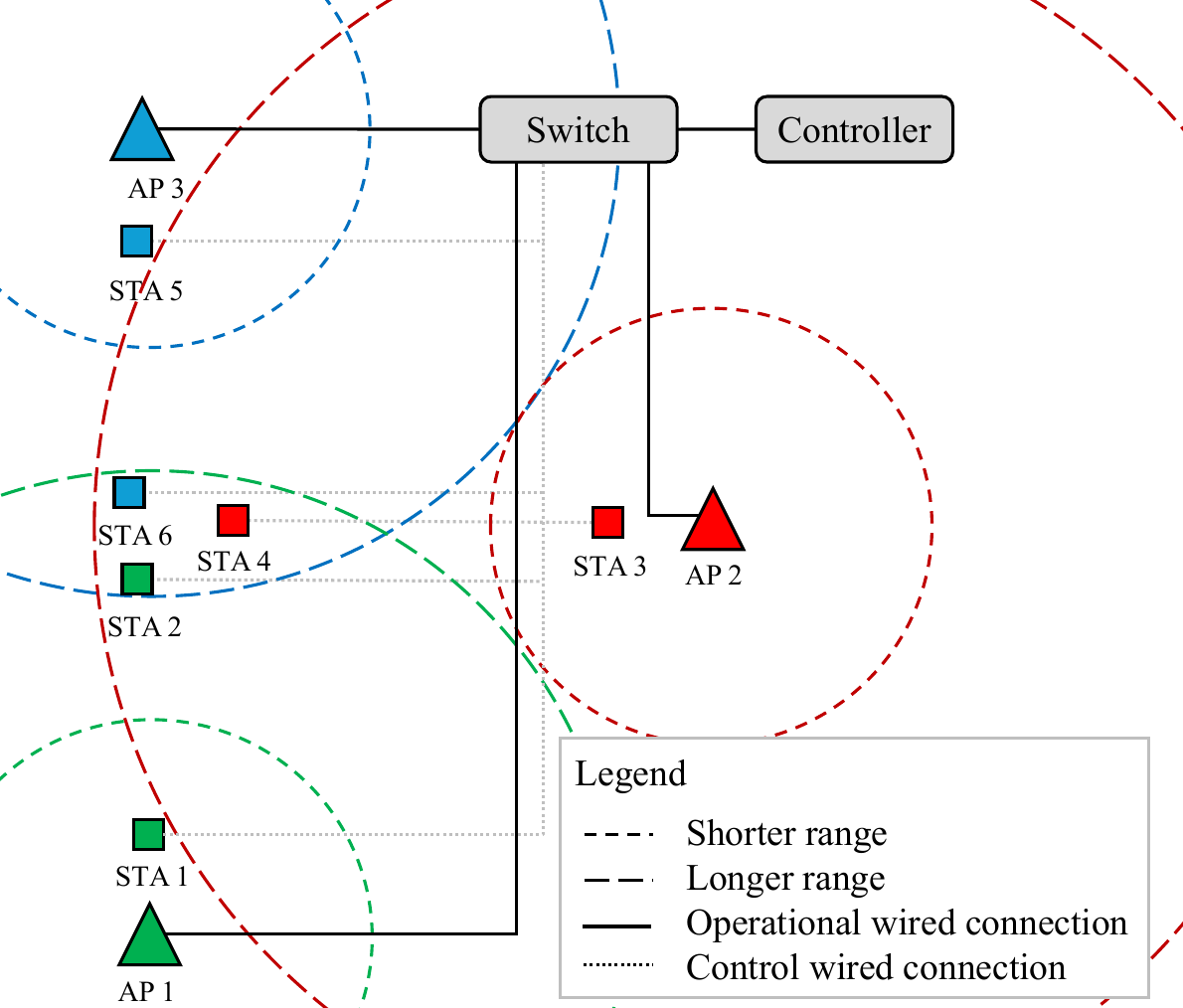}
\caption{Schematic illustration of the testbed topology. 
}
\label{fig:testbed}
\end{figure}

\subsection{Experimental Setup}
\begin{figure}[!t]
\centering
\includegraphics[width=0.95\columnwidth]{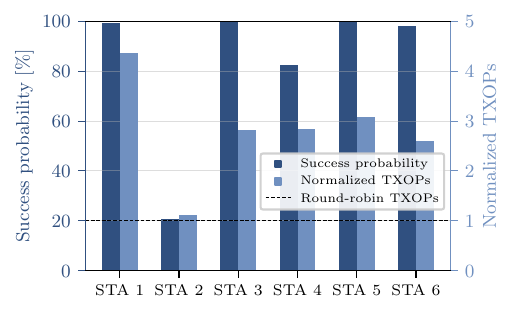}
\caption{Testbed results: success probability under single concurrent transmissions and number of TXOPs given by H-MABs to each station normalized to a perfect round robin scheme (i.e., the long term performance of DCF).}
\label{fig:testbed_ps}
\end{figure}

The network topology comprises three APs and six stations (Fig.~\ref{fig:testbed}). All devices use IEEE 802.11n and operate on the same \SI{20}{MHz} channel centered at \SI{3.686}{GHz} (to avoid external interference). Each AP serves two stations: one placed near its serving AP (\SI{1}{\meter} away) and the other in the boundary coverage zone between all APs (\SI{6}{\meter} to \SI{8}{\meter} away from its AP). However, due to different station placement (i.e., different propagation and channel conditions) their transmission success probability varies under a given MCS (Fig.~\ref{fig:testbed_ps}, left Y axis, shows the success probability of different stations for the MCS index set to 6) which constitutes a complex real scenario.

In this setting, concurrent C-SR transmissions are possible by adjusting the transmit power levels of APs. 
Three discrete power levels are allowed in the testbed: the default value (\SI{17}{dBm} for openwifi APs), the default decreased by \SI{6}{dB} and the default decreased by \SI{12}{dB}. All APs are interconnected to a switch and further to a network controller via a wired backbone.
This controller collects rewards and determines C-SR scheduling  according to Section~\ref{sec:mabs}.

We configure openwifi so that parallel transmissions can start synchronously. A transmission slot length of \SI{307}{\micro\second}  restricts the time slot duration to exactly one \SI{1000}{B} frame transmitted with MCS~6 (Table~\ref{tab_testbed_settings}). 
We increase the CCA threshold significantly to force nodes to perform transmissions. We update the configuration selected by the agent every 10 frame transmissions; we call this a \textit{step}. 
To accurately measure the number of received frames, we implement a traffic generation and monitoring framework, which reports the number of sent and received frames over the wired link. 

\begin{table}
\caption{Testbed parameter settings}
\label{tab_testbed_settings}
\centering
\begin{tabular}{@{}ll@{}}
\toprule
\textbf{Parameter}         & \textbf{Value}            \\ \midrule
Band                       & \SI{3.7}{GHz}                   \\
PHY                        & IEEE   802.11n           \\
Channel   width            & \SI{20}{MHz}                  \\
Spatial   streams          & 1, SISO                   \\
MCS index                       & 6                  \\
Tx power attenuation levels  & $\{0, -6, -12\}$ \SI{}{dB}         \\
Frame   size               & \SI{1000}{B}                  \\ 
CCA threshold              & \SI{-52}{dBm}                  \\ 
Transmission slot length   & \SI{307}{\micro\second}       \\ 
Packets per slot  & 1       \\ 
Slots per configuration  & 10       \\  
Number of repetitions     & 10                        \\
\bottomrule
\end{tabular}
\end{table}

\begin{figure}[!t]
\centering
\includegraphics[width=0.95\columnwidth]{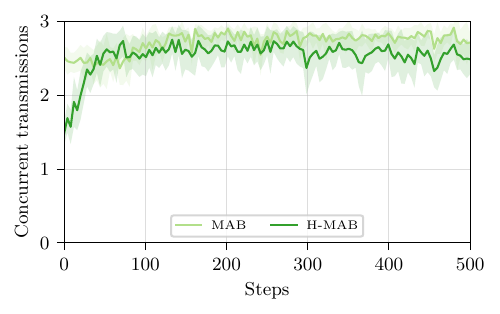}
\caption{Comparison of MAB and H-MAB C-SR operation in the testbed: average number of concurrent transmissions.}
\label{fig:testbed_results}
\end{figure}

\begin{figure}[!t]
\centering
\includegraphics[width=0.95\columnwidth]{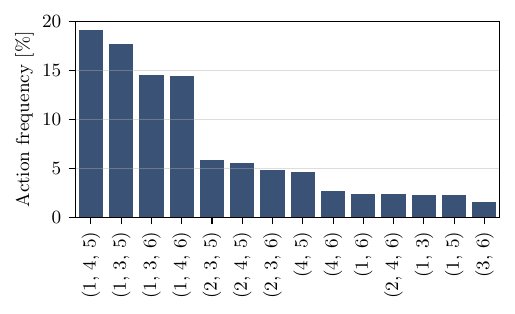}
\caption{Actions taken by H-MAB in the last 200 steps of the experiments. The X axis displays the set of stations selected for parallel transmissions. Only actions with a frequency above 0.5\% are shown.}
\label{fig:testbed_actions}
\end{figure}

\subsection{Results}
Our results show that, in this C-SR setting, both flat MABs and H-MABs quickly find appropriate AP-station pairs and transmit power settings that enable three concurrent transmissions (Fig.~\ref{fig:testbed_results}). Since the tested topology is relatively small, 
flat MABs slightly outperform H-MABs. Nonetheless, the results validate that the proposed C-SR approach based on MABs is feasible and works well in a real environment.

We also show actions taken by H-MABs gathered from the last 200 steps of all runs (Fig.~\ref{fig:testbed_actions}). More than 80\% of actions taken involve three parallel transmissions. As expected, C-SR configurations including station 2 are less frequent because they result in lower throughput.
To explain this, we examine the underlying success probability of transmissions to each station (Fig.~\ref{fig:testbed_ps}, left Y axis), where station~2 has noticeably worse radio conditions.
As shown in Fig.~\ref{fig:testbed_ps} (right Y axis), station 2 preserves its TXOPs as compared to DCF.
This behavior confirms that H-MABs not only configure parallel transmissions but also adjust to the observed environmental conditions. 
Furthermore, this confirms that fairness among stations is not a straightforward goal for C-SR, as some stations may be unable to benefit from C-SR (but should be granted no fewer TXOPs than under DCF, which we do achieve with the proposed framework).

\section{Discussion and Future Work}
\label{sec:conclusions}

Our first contribution is the theoretical upper bound model of C-SR. We show that it can be optimized for fairness or throughput and provide an additional scalability analysis (Appendix~\ref{appendix}). We expect new C-SR scheduling solutions to emerge as work on IEEE 802.11bn progresses, and our model provides the research community with an upper bound baseline. As future work in this area, we plan to address other optimization goals (such as proportional fairness or assigning each station a fair number of transmission opportunities, ideally no less than under DCF) and add lexicographic optimization. With these updates, the model should provide more realistic (less extreme) results than those presented here.

Our second contribution, inspired by the probing-based algorithms for data rate selection algorithms in Wi-Fi networks, is the application of MABs (both flat and hierarchical) for solving the C-SR scheduling problem.
Our results show that MABs are a highly promising learning mechanism. 
In addition to good performance (high throughput, low convergence time for small and medium-sized networks) and resilience to abrupt topology changes, an important advantage is their simplicity, which makes the solution easy to deploy in hardware. For large-scale networks, we evaluate clustered MABs, which also demonstrate strong performance and rapid convergence. Furthermore, we assess the impact of legacy APs on the operation of our proposal and find that H-MABs are reasonably resilient.

A classical MAB is defined for a constant number of arms (choices), and this is in contrast to the dynamic behavior of users who join and leave the network. Although this may seem a blocking characteristic, we argue that this limitation is tied only to the initial implementation presented. Consider a highly realistic scenario where APs operate with non-full buffers while stations arrive and depart freely. In these cases, it is reasonable to preallocate
agents up to some maximal number of APs and stations, respectively. Some of the preallocated agents or arms are not used and are called coldspares. When new stations arrive, we can allocate their adequate coldspares. When APs have empty buffers, we can move them to coldspares.
The update operation remains unchanged, as we update only the agents that are in use. We modify the sampling to ensure that the coldspare is never selected. For this, we propose rejection sampling that repeats the sample operation when the coldspare is selected. This procedure fishes for probabilistic agents, and we have an undergoing effort to implement this dynamic based on a proof-of-concept implementation that can be found in \emph{Reinforced-lib}~\cite{wojnar2023reinforced}.

Furthermore, we foresee the following other future research items. Although the presented MABs use aggregate network throughput as a reward, other optimization goals (e.g., latency or fairness) can be considered. To simplify the presentation, we have idealized the MCS selection, but it can and should be jointly optimized along with the transmission pairs and transmit power. 
Our H-MAB framework can operate in a distributed manner (having independent agents on different APs and synchronizing updates), but this would complicate the signaling and introduce additional overhead, so we do not describe it in this work. However, a distributed learning scheme with independent agents (i.e., assuming little or no communication) would be interesting to study (especially to see if convergence can be achieved). Finally, large-scale scenarios with increased mobility, external interference and other dynamics, as well as uplink traffic should be considered in the future. 


%

\appendices
\section{Scalability Analysis of the Optimization Model}
\label{appendix}

The practical applicability of the proposed optimization model (Section~\ref{sec:model}) hinges on its ability to efficiently handle real-world network scenarios, which often involve a larger number of APs and stations. Therefore, it is crucial to assess the scalability of the model, i.e., how its computational requirements grow as the network size increases. This analysis will provide valuable insights into the model's feasibility for deployment in real-world sized Wi-Fi networks.


The performance and scalability of the optimization model are significantly influenced by the underlying solver used to find the optimal solution. In this study, we compare two solvers: COIN-OR branch and cut (CBC) \cite{cbc} and IBM ILOG CPLEX Optimization Studio (CPLEX) \cite{cplex}.
CBC is an open-source MILP solver readily available in the PuLP library in Python. Its accessibility and ease of integration make it a popular choice for prototyping and experimentation. However, being a general-purpose solver, it is not optimized for the specific structure of our C-SR scheduling problem.
CPLEX is a commercial solver widely recognized for its high performance and advanced algorithms tailored for solving large-scale linear and MILP problems. While it requires a separate license, 
its computational efficiency and scalability make it a strong candidate for handling complex optimization tasks, such as C-SR scheduling.


\begin{figure}[!t]
\centering
\includegraphics[width=0.95\columnwidth]{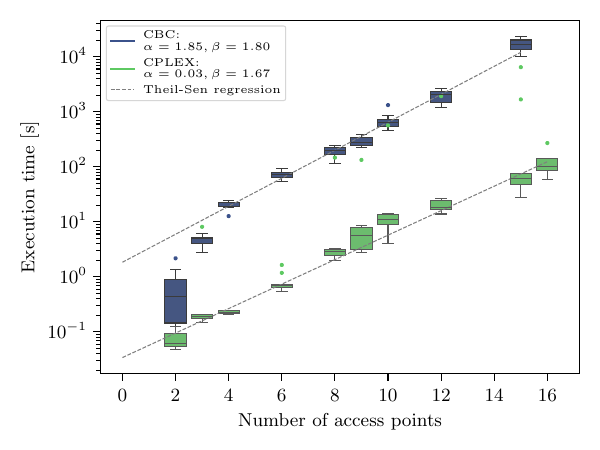}
\caption{Scalability of the optimization solution depending on the solver used.}
\label{fig:scalability}
\end{figure}

We assess the scalability of the T-Optimal model in the multi-room topology. We vary the number of APs within this grid to simulate networks of increasing size. The execution time of both solvers (CBC and CPLEX) is measured for each network size 10 times, providing a direct comparison of their scalability.

Fig.~\ref{fig:scalability} presents the execution time of both solvers on a logarithmic scale as a function of the number of APs in the network. The Theil-Sen regression lines \cite{theil-sen}, chosen for their robustness to outliers, are fitted for both solvers in log space to determine the quantitative measure of the scalability of each solver. The results clearly indicate that the execution time for both solvers increases exponentially with network size, as expected. The regression lines capture this exponential relationship, with their slopes representing the exponents ($\beta$) of the time complexity. CPLEX demonstrates superior scalability ($\beta = 1.67$) compared to CBC ($\beta = 1.80$), which implies that CPLEX's execution time grows at a slower rate as the network size increases, making it computationally more efficient, especially for larger networks.

While CPLEX generally exhibits lower execution times, it also shows greater variation, with some outliers having execution times comparable to those of the CBC solver. This suggests that CPLEX's performance may be more sensitive to specific problem instances or network configurations, whereas CBC's performance tends to be more consistent, albeit slower overall.

\section*{Acknowledgments}

This research was funded by the National Science Centre, Poland (2020/39/I/ST7/01457) and by the German Research Foundation (DFG DR 639/28-1). 
This research was also partially funded by the Flemish FWO SBO S003921N VERI-END.com project.
Additionally, the work of Boris Bellalta was supported by Wi-XR PID2021123995NB-I00 (MCIU/AEI/FEDER,UE), and the ICREA Academia program.
We gratefully acknowledge the computing infrastructure provided with support from the National Research Institute, grant number POIR.04.02.00-00-D008/20-01 on ``National Laboratory for Advanced 5G Research'' (PL-5G) as part of Measure 4.2 Development of modern research infrastructure of the science sector 2014-2020 financed by the European Regional Development Fund. 

\ifCLASSOPTIONcaptionsoff
  \newpage
\fi



%

\printbibliography

%
\begin{IEEEbiography}[{\includegraphics[width=1in,height=1.25in,clip,keepaspectratio]{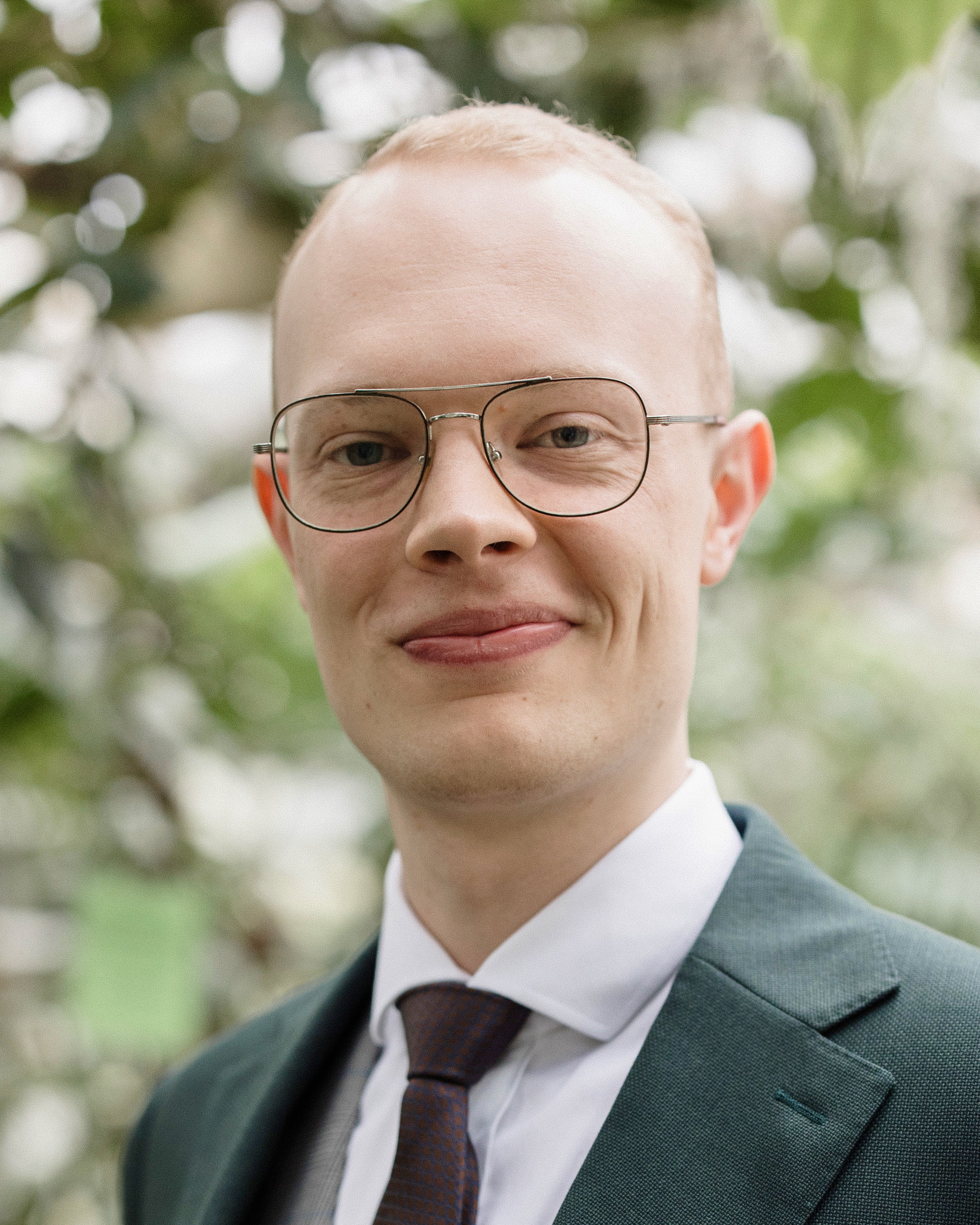}}]{Maksymilian Wojnar} is a PhD candidate and researcher in machine learning and wireless networks, with an MSc degree in computer science from the AGH University of Krakow (2024). His research focuses primarily on optimizing wireless networks through machine learning techniques, particularly reinforcement learning. He is actively involved in collaborative research projects within these areas. Additionally, his work extends to the broader application of machine learning, with contributions in generative neural networks.
\end{IEEEbiography}

\begin{IEEEbiography}[{\includegraphics[width=1in,height=1.25in,clip,keepaspectratio]{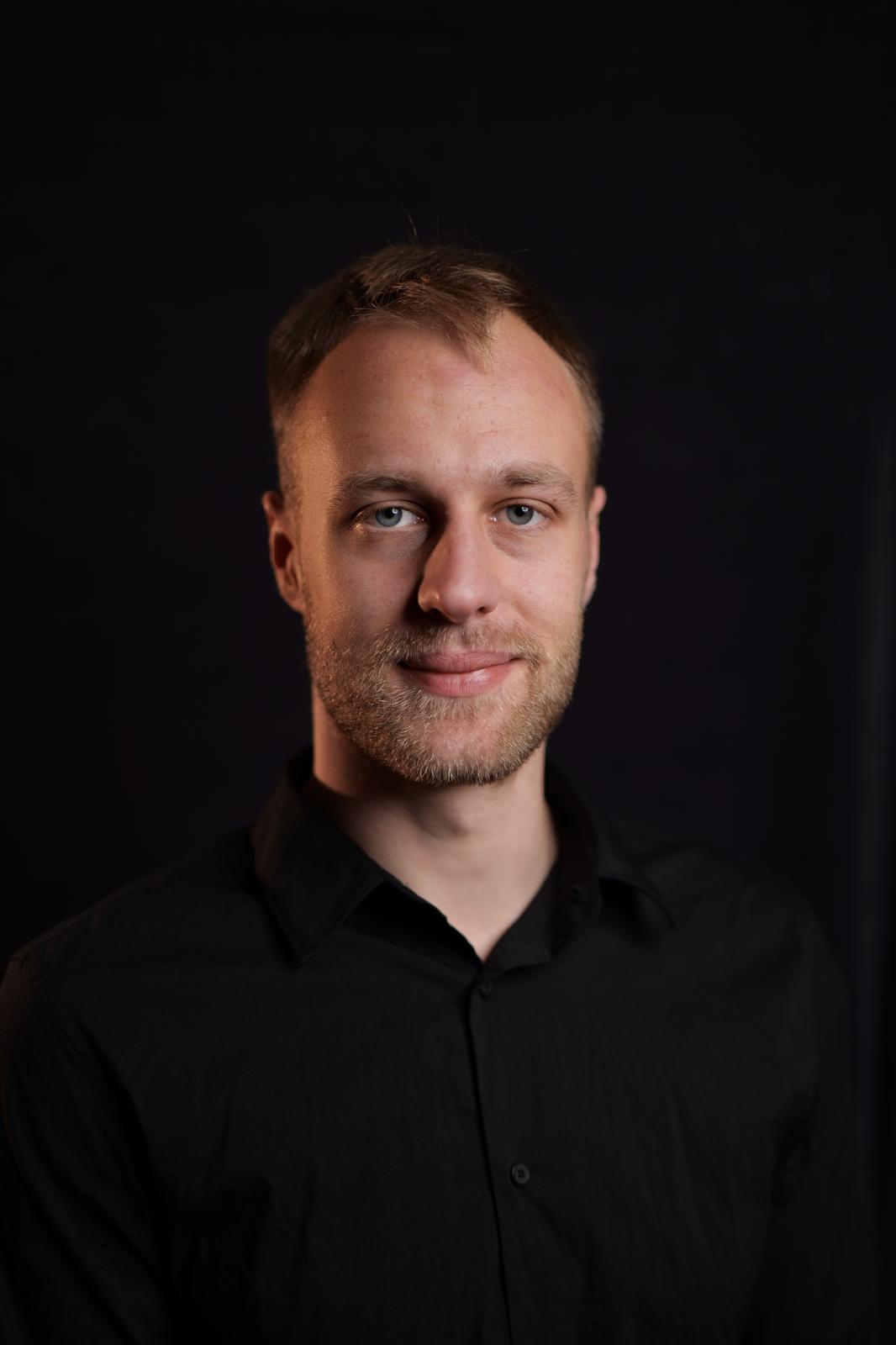}}]{Wojciech Ciężobka} received his MSc degree (with honors) in Data Science from the AGH University of Krakow in 2024, where his research focused on applying reinforcement learning to optimize future Wi-Fi networks. The work presented in this article stems from his time as a Machine Learning Researcher at AGH. He now applies his expertise in machine learning and optimization in the logistics industry, where he coordinates the development and deployment of a new generation of freight forwarding software.
\end{IEEEbiography}

\begin{IEEEbiography}[{\includegraphics[width=1in,height=1.25in,clip,keepaspectratio]{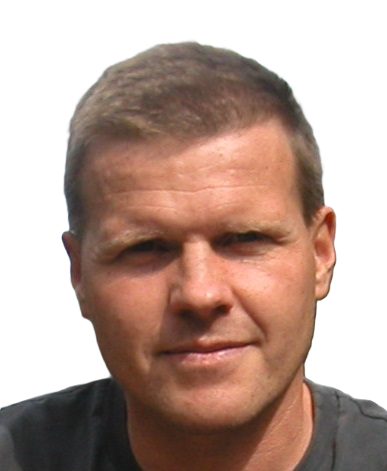}}]{Artur Tomaszewski} received the MSc and PhD degrees in telecommunications, in 1990 and 1993, respectively, and in 2014 received the DSc degree (habilitation) in technical sciences, all from the Warsaw University of Technology (WUT). He was the head of the Computer Networks and Services Division at WUT. Currently, he is a professor at the Faculty of Electronics, Telecommunications and Informatics, Gdańsk University of Technology (GUT). His research areas are communications networks architecture, network control and management, and network design and optimization.
\end{IEEEbiography}

\begin{IEEEbiography}[{\includegraphics[width=1in,height=1.25in,clip,keepaspectratio]{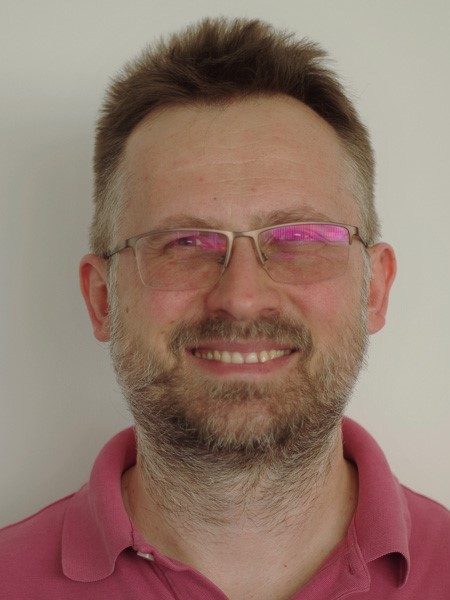}}]{Piotr Chołda} is an associate professor at the Institute of Telecommunications, AGH University of Krakow, Poland. He earned his habilitation 
in Telecommunications from AGH University in 2016. His research interests include the management of communication networks and computer systems, with a particular focus on risk-aware performance evaluation and design, resource optimization, and control using machine learning methods, as well as novel approaches in cryptology. Dr.~Chołda has participated in several European research projects, including NOBEL, 
Euro-NF, SmoothIT, STAR, RECODIS,
and PQ-REACT.
\end{IEEEbiography}

\begin{IEEEbiography}[{\includegraphics[width=1in,height=1.25in,clip,keepaspectratio]{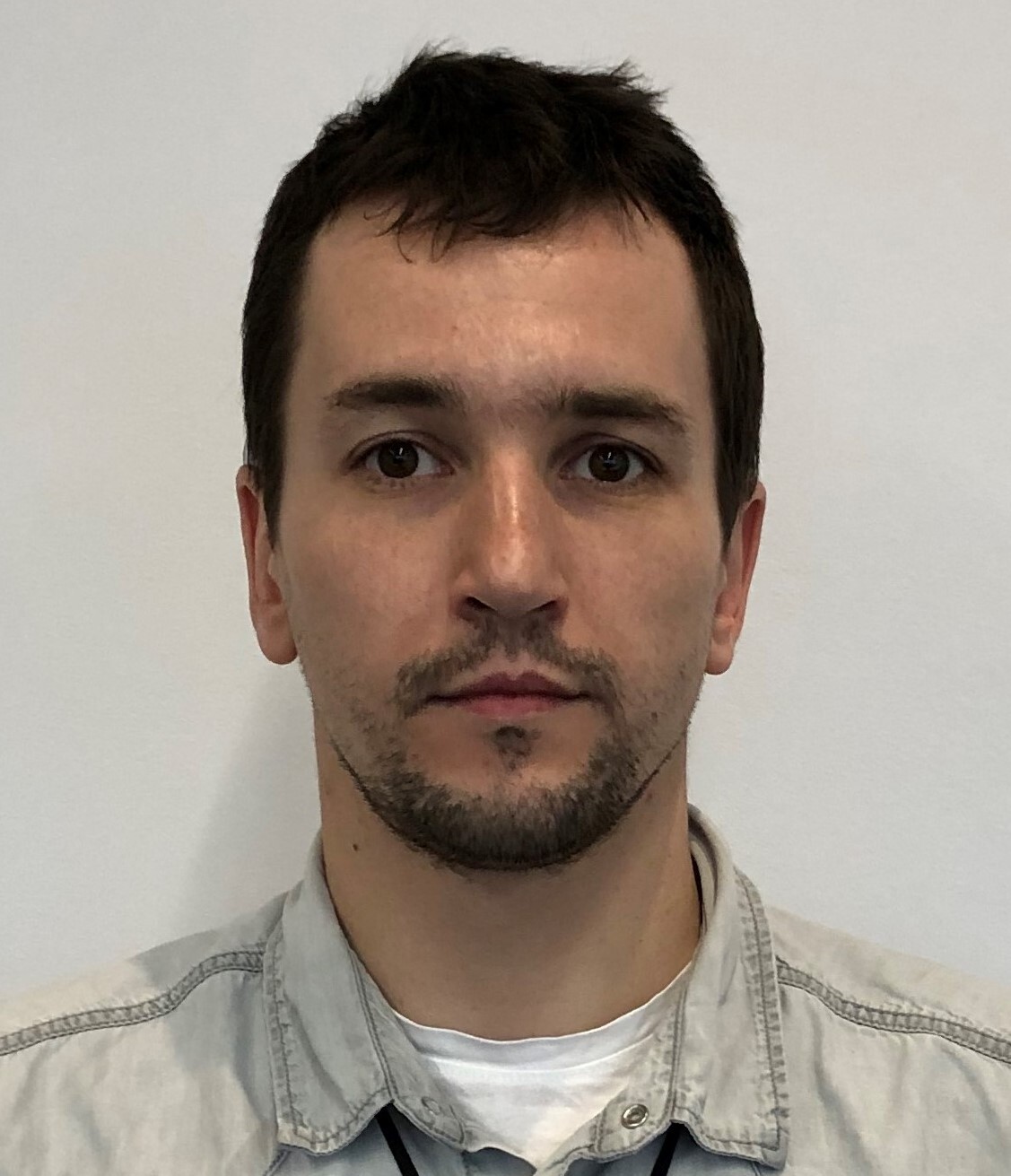}}]{Krzysztof Rusek} is an assistant professor at AGH University and a data scientist at the Barcelona Neural Networking Center. He defended his PhD thesis on queuing theory in 2016 at AGH.
His main research interests are performance evaluation of telecommunications systems, machine learning and data mining. 
Currently, he is working on the application of graph neural networks and probabilistic modeling for performance evaluation of communications systems.
\end{IEEEbiography}

\begin{IEEEbiography}[{\includegraphics[width=1in,height=1.25in,clip,keepaspectratio]{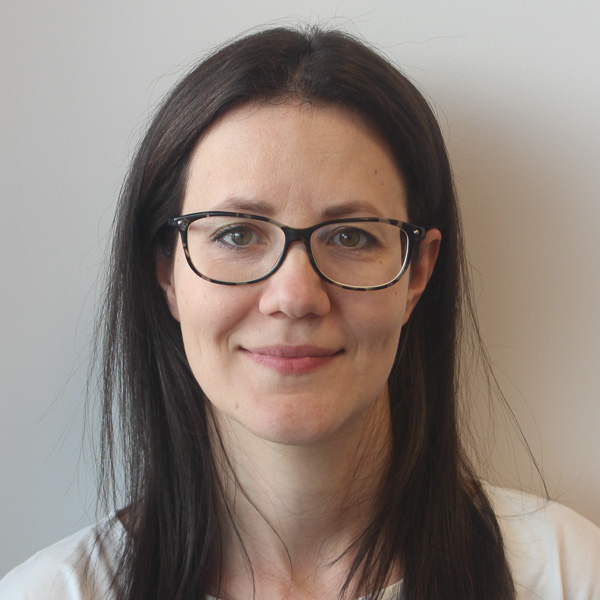}}]{Katarzyna Kosek-Szott} received her MSc and PhD degrees in telecommunications (both with honors) from the AGH University of Krakow, Poland in 2006 and 2011, respectively. In 2016 she received her habilitation degree and in 2023 -- a full professor title.  Currently she is working as a full professor at the Institute of Telecommunications, AGH University of Krakow.  Her general research interests are focused on wireless networking. The major topics include wireless LANs, quality of service provisioning, novel amendments to the IEEE 802.11 standard, performance improvement with machine learning, and coexistence of different radio technologies in unlicensed bands. She is a reviewer for international journals and conferences. She has been involved in several European projects: DAIDALOS II, CONTENT, CARMEN, FLAVIA, PROACTIVE, RESCUE as well as grants supported by the Polish Ministry of Science and Higher Education and the National Science Centre. In 2015-2020, she served as a member of the editorial board of the Ad Hoc Networks journal published by Elsevier. In 2018, she served as an European Research Council Advanced Grant Panel member for the Systems and Communication Engineering area. In 2020, she served as an European Research Council Consolidator Grant Panel member for the Systems and Communication Engineering area.
\end{IEEEbiography}

\begin{IEEEbiography}
[{\includegraphics[width=1in,height=1.25in,clip,keepaspectratio]{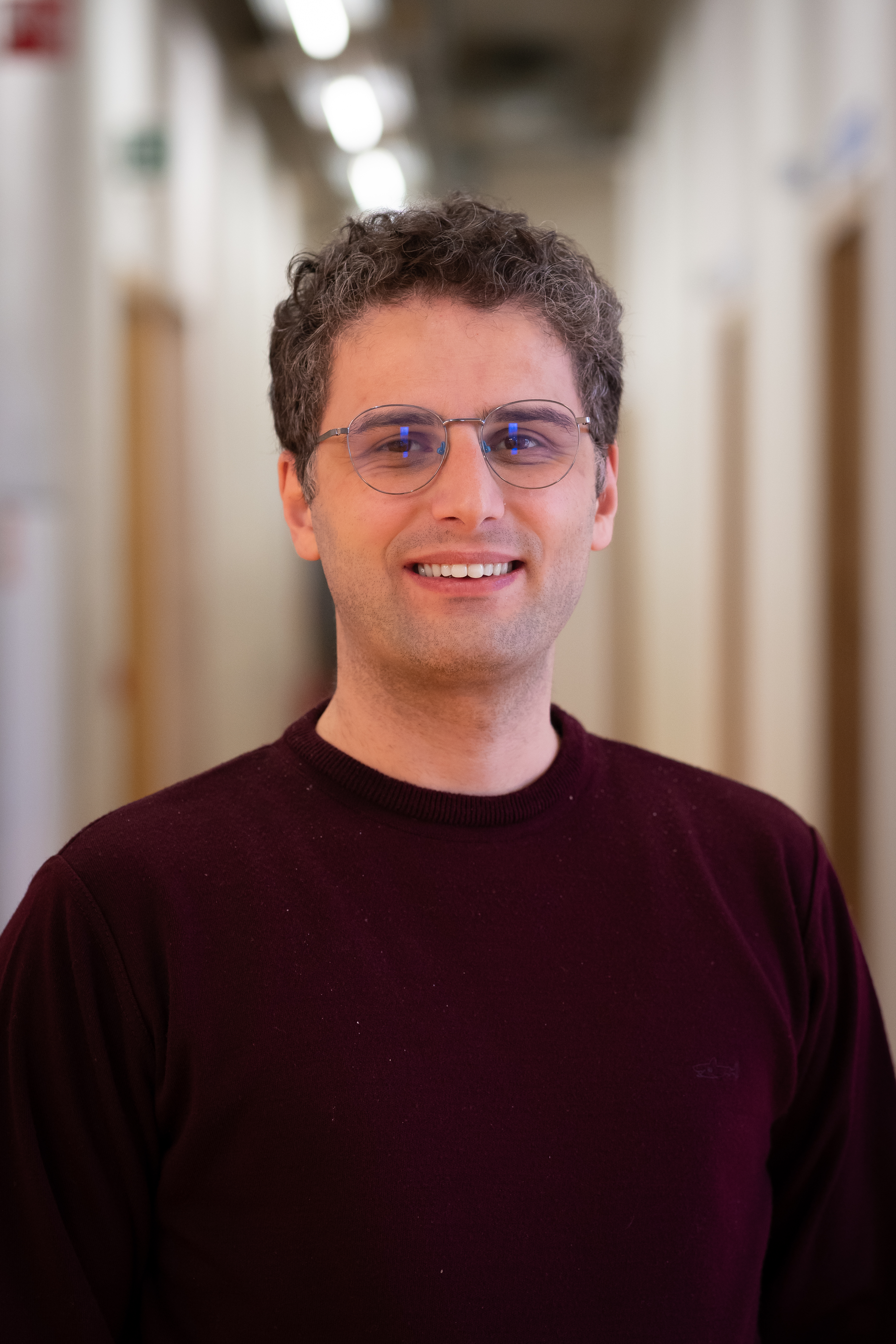}}]
{Jetmir Haxhibeqiri} (IEEE member) received the Masters degree in Engineering (information technology and computer engineering) from RWTH Aachen University, Germany (2013). In 2019, he obtained a PhD in Engineering Computer Science from Ghent University with his research on flexible and scalable wireless communication solutions for industrial warehouses and logistics applications. Currently, he is a senior researcher in the Internet Technology and Data Science Lab (IDLab) of Ghent University and imec. His current research interests include wireless communications technologies (IEEE 802.11, IEEE 802.15.4e, LoRa) and their application, IoT, wireless time sensitive networking, in-band network monitoring and wireless network management.
\end{IEEEbiography}

\begin{IEEEbiography}[{\includegraphics[width=1in,height=1.25in,clip,keepaspectratio]{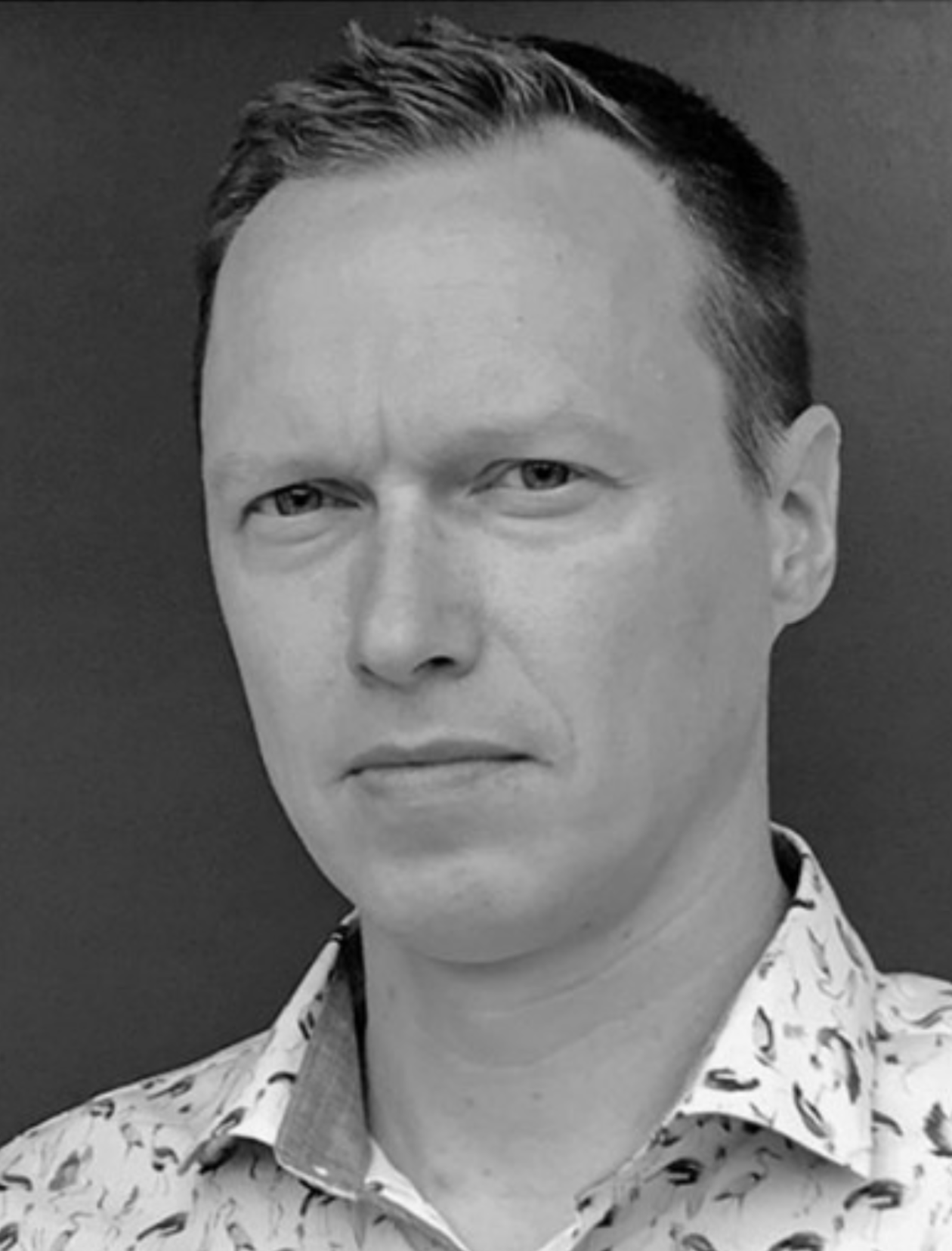}}]
{Jeroen Hoebeke} received the master's degree in engineering computer science from Ghent University, in 2002, and the PhD degree in engineering computer science, in 2007, with his research on adaptive ad hoc routing and virtual private ad hoc networks. He is currently an Associate Professor with the Internet Technology and Data Science Laboratory, Ghent University and imec. He is the author or co-author of more than 200 publications in international journals or conference proceedings. He is conducting and coordinating research on wireless (IoT) connectivity, embedded communication stacks, deterministic wireless communication, and wireless network management.   
\end{IEEEbiography}

\begin{IEEEbiography}[{\includegraphics[width=1in,height=1.25in,clip,keepaspectratio]{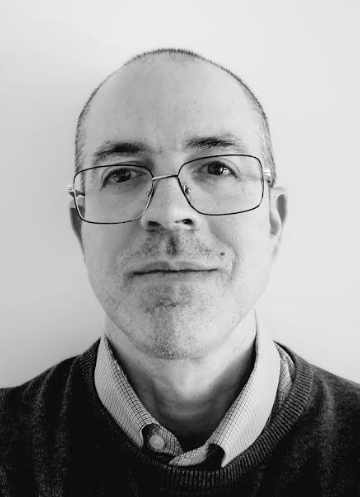}}]
{Boris Bellalta} (IEEE Senior Member) is a full professor at Universitat Pompeu Fabra (UPF), where he leads the Wireless Networking group. His research interests lie in the area of wireless networks and performance evaluation, with a particular emphasis on Wi-Fi technologies and machine learning-based adaptive systems. He has supervised more than 15 PhD students. In 2024, he was awarded the ICREA Academia fellowship by the Catalan Government.  
\end{IEEEbiography}

\begin{IEEEbiography}[{\includegraphics[width=1in,height=1.25in,clip,keepaspectratio]{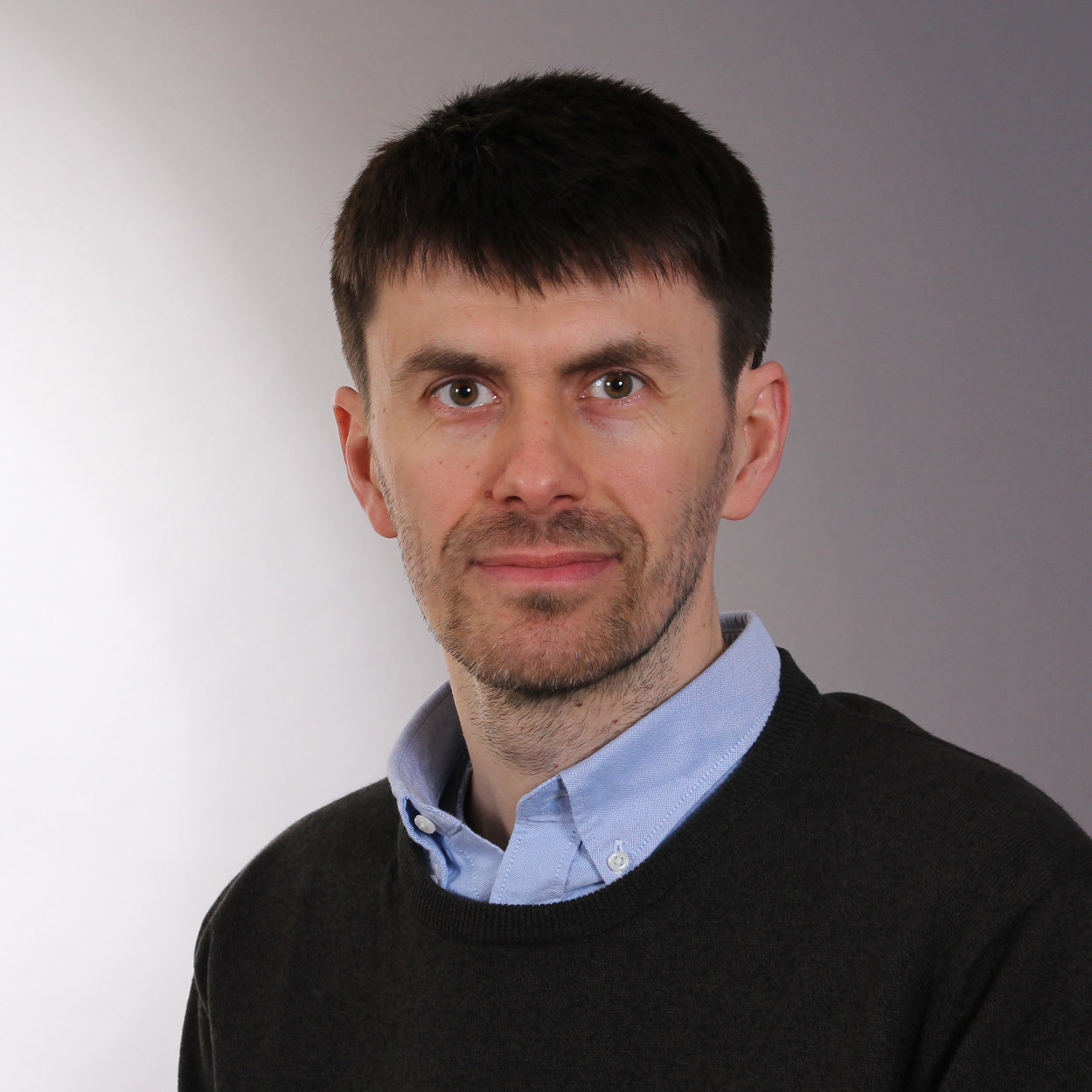}}]{Anatolij Zubow} (IEEE Senior Member) received the MSc and PhD degrees from the Department of Computer Science, Humboldt Universität zu Berlin, in 2004 and 2009, respectively. He is a senior researcher at the department of Electrical Engineering and Computer Science of Technische Universität Berlin. His research interest is on emerging wireless network architectures. Recently he focuses mainly on coexistence of heterogeneous wireless technologies in unlicensed spectrum and high-performance WiFi networks. He has strong interest in prototyping, experimental work and testbeds. In the past, Dr. Zubow had multiple research visits undertaken at the NEC Network Laboratories in Heidelberg, where he was working on future mobile networks.
\end{IEEEbiography}

\begin{IEEEbiography}[{\includegraphics[width=1in,height=1.25in,clip,keepaspectratio]{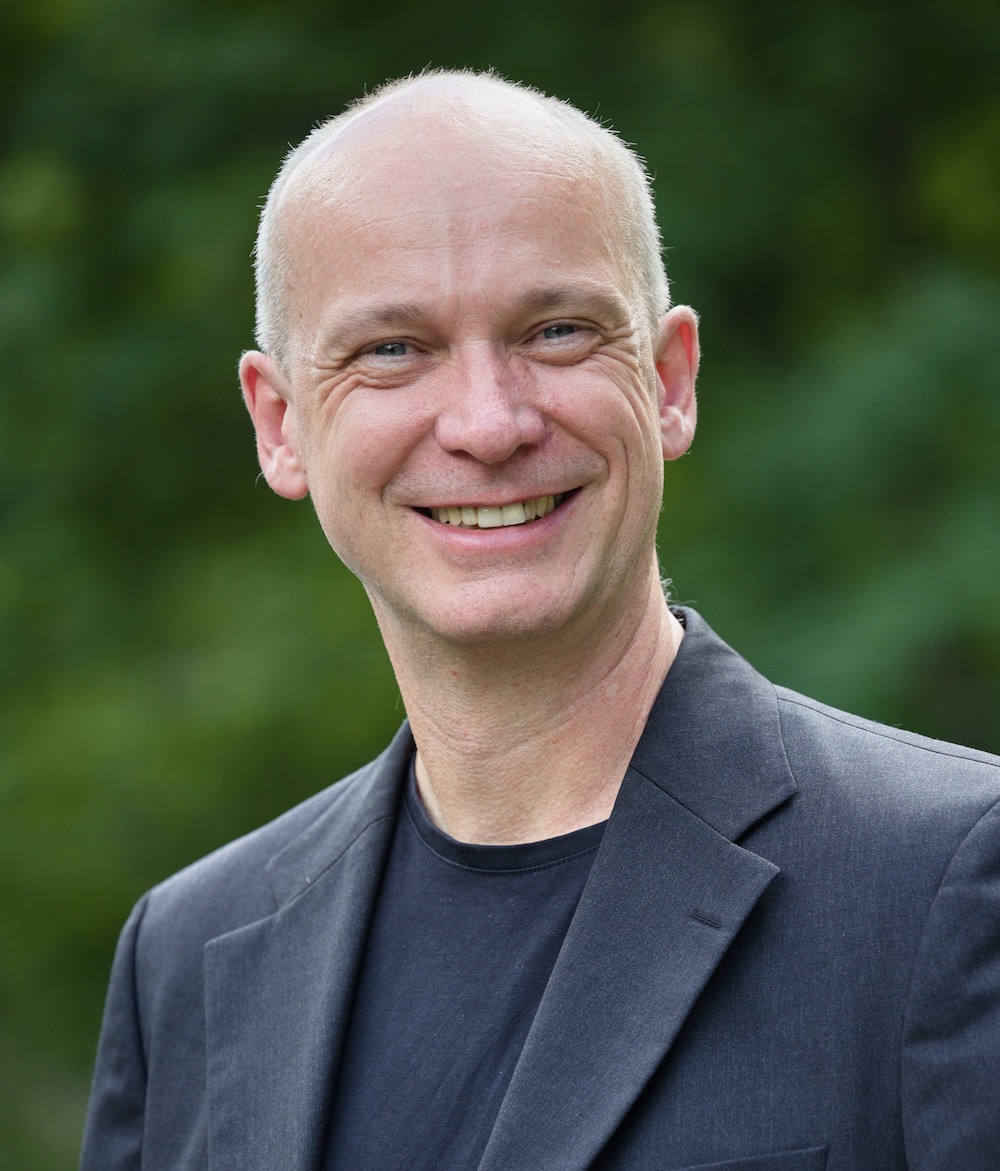}}]{Falko Dressler} (IEEE Fellow) is full professor and Chair for Telecommunication Networks at the School of Electrical Engineering and Computer Science, TU Berlin. He received his MSc and PhD degrees from the Dept. of Computer Science, University of Erlangen in 1998 and 2003, respectively.
Dr. Dressler has been associate editor-in-chief for IEEE Trans. on Network Science and Engineering, IEEE Trans. on Mobile Computing and Elsevier Computer Communications as well as an editor for journals such as IEEE/ACM Trans. on Networking, Elsevier Ad Hoc Networks, and Elsevier Nano Communication Networks. He has been chairing conferences such as IEEE INFOCOM, ACM MobiSys, ACM MobiHoc, IEEE VNC, IEEE GLOBECOM. He authored the textbooks Self-Organization in Sensor and Actor Networks published by Wiley \& Sons and Vehicular Networking published by Cambridge University Press. He has been an IEEE Distinguished Lecturer as well as an ACM Distinguished Speaker.
Dr. Dressler is an IEEE Fellow, an ACM Fellow, and an AAIA Fellow. He is a member of the German National Academy of Science and Engineering (acatech). He has been serving on the IEEE COMSOC Conference Council and the ACM SIGMOBILE Executive Committee. His research objectives include next generation wireless communication systems in combination with distributed machine learning and edge computing for improved resiliency. Application domains include the internet of things, cyber-physical systems, and the internet of bio-nano-things.
\end{IEEEbiography}

\begin{IEEEbiography}[{\includegraphics[width=1in,height=1.25in,clip,keepaspectratio]{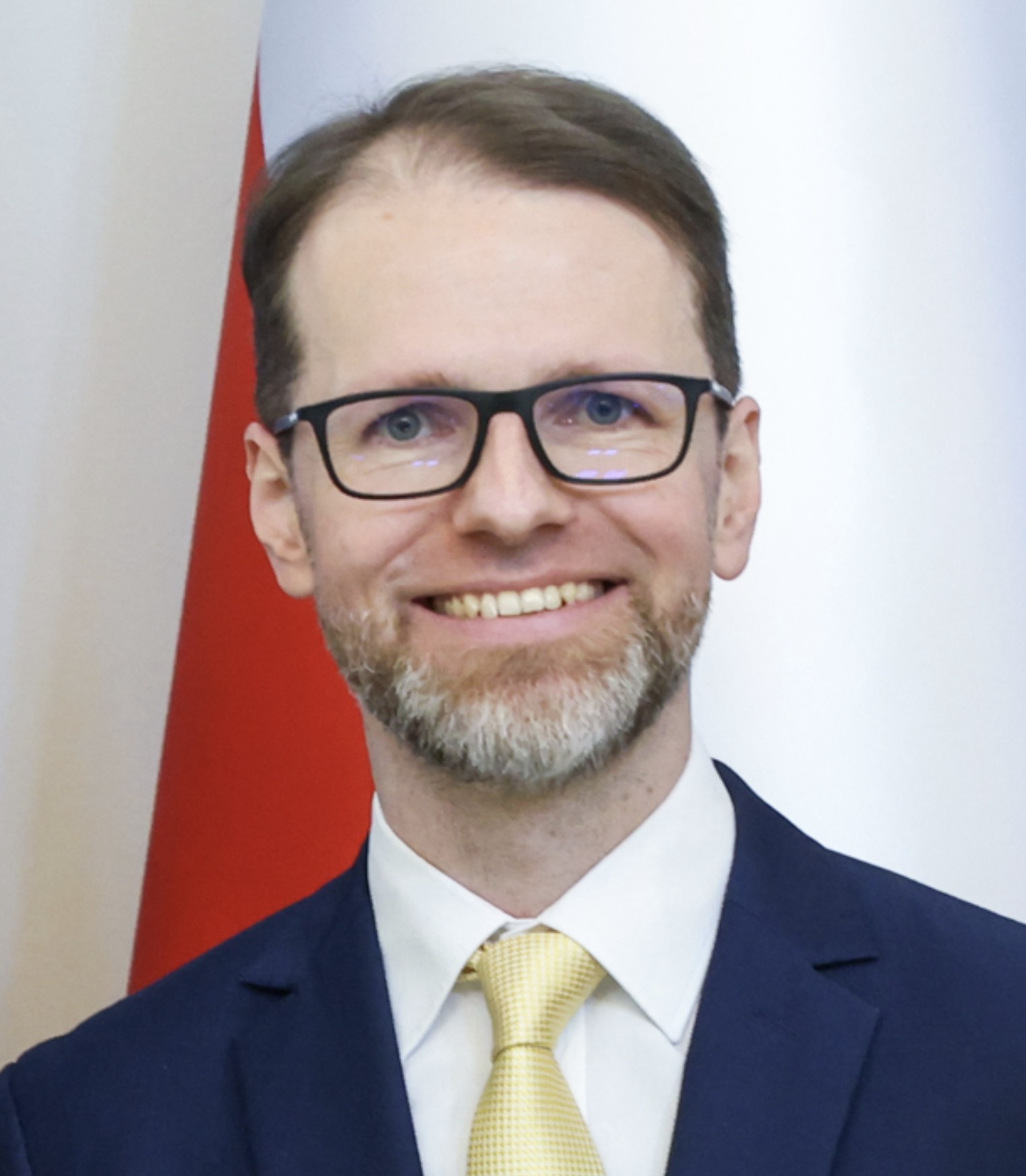}}]{Szymon Szott} received his MSc and PhD degrees in telecommunications (both with honors) from the AGH University of Krakow, Poland in 2006 and 2011, respectively. In 2025, he became a full professor at the Institute of Telecommunications of AGH University. In 2013, he was a visiting researcher at the University of Palermo (Italy) and at Stanford University (USA). His professional interests are related to wireless local area networks (channel access, quality of service, security, inter-technology coexistence). He is a member of the IEEE 802.11 Working Group. In the past, he has been a member of ETSI's Network Technology Working Group Evolution of Management towards Autonomic Future Internet (AFI), a senior member of IEEE, and on the management board of the Association of Top 500 Innovators. He has been involved in several European projects (DAIDALOS II, CONTENT, CARMEN, MEDUSA, FLAVIA, PROACTIVE, RESCUE, MLDR) as well as grants supported by the Ministry of Science and Higher Education and the National Science Centre. 
\end{IEEEbiography}







\end{document}